\documentclass[10pt]{iopart}
\usepackage[dvipsnames,table]{xcolor}
\usepackage{graphicx,subfig,caption}
\usepackage{hyperref}
\usepackage{cleveref}
\usepackage{multirow}
\usepackage{dblfloatfix}
\usepackage[backend=bibtex,sorting=none,style=numeric-comp]{biblatex}
 \usepackage{indentfirst}

\bibliography{BIB}

\begin{document}
\title[Computational study of positive and negative streamers in twin SDBD via 2D PIC simulations]{Computational study of simultaneous positive and negative streamer propagation in a twin surface dielectric barrier discharge via 2D PIC simulations}

\author{Quan-Zhi Zhang$^{1,*}$, R T Nguyen-Smith$^{2,*}$, 
F Beckfeld$^2$, Yue Liu$^3$, T Mussenbrock$^3$, P Awakowicz$^2$ and J Schulze$^{1,2}$}

\address{$^1$ School of Physics, Dalian University of Technology, Dailan 16024, P.R. China}
\address{$^2$ Chair of Electrical Engineering and Plasma Technology, Faculty of Electrical Engineering and Information Sciences, Ruhr University Bochum, Germany}
\address{$^3$ Chair of Plasma Technology, Faculty of Electrical Engineering and Information Sciences, Ruhr University Bochum, Germany}
$^*$ Authors contributed equally to this work
\ead{\mailto{qzzhangdlut@gmail.com}, \mailto{rtnguyensmith@gmail.com}}

\begin{abstract}
    
    The propagation mechanisms of plasma streamers have been observed and investigated in a surface dielectric barrier discharge (SDBD) using 2D particle in cell simulations. The investigations are carried out under a simulated air mixture, 80\% N$_2$ and 20\% O$_2$, at atmospheric pressure, 100$\,$kPa, under both DC conditions and a pulsed DC waveform that represent AC conditions. The simulated geometry is a simplification of the symmetric and fully exposed SDBD resulting in the simultaneous ignition of both positive and negative streamers on either side of the Al$_2$O$_3$ dielectric barrier. In order to determine the interactivity of the two streamers, the propagation behavior for the positive and negative streamers are investigated both independently and simultaneously under identical constant voltage conditions. An additional focus is implored under a fast sub nanosecond rise time square voltage pulse alternating between positive and negative voltage conditions, thus providing insight into the dynamics of the streamers under alternating polarity switches. It is shown that the simultaneous ignition of both streamers, as well as using the pulsed DC conditions, provides both an enhanced discharge and an increased surface coverage. It is also shown that additional streamer branching may occur in a cross section that is difficult to experimentally observe. The enhanced discharge and surface coverage may be beneficial to many applications such as, but are not limited to: air purification, volatile organic compound removal, and plasma enhanced catalysis.
    
\end{abstract}

\submitto{\PSST}
\maketitle
\ioptwocol
%main document
\section{Introduction}
    Dielectric barrier discharges (DBDs) are plasma discharges incorporating at least one layer of dielectric material separating the two electrodes. The dielectric barrier limits the charge transfer and thus the current flow typically producing a non thermal plasma at atmospheric conditions. This non thermal nature allows for the efficient generation of reactive species thereby providing multiple possibilities in biomedical, surface, and industrial applications \cite{Brandenburg2017,HHKim2004}. DBDs are classifiable into two main categorical descriptors: volumetric and surface DBDs. Volume dielectric barrier discharges (VDBDs) are classifiable from DBDs by having a gas gap and a dielectric barrier present between the two electrodes, producing either homogeneous or filamantary like plasmas depending on the conditions \cite{Kogelschatz2010}. Surface dielectric barrier discharges (SDBDs) on the other hand, have only the dielectric layer directly separating the two electrodes; a plasma is thereby only able to ignite along the surface of the dielectric. Due to the possibility of having a thin structure, SDBDs may have particularly low flow resistance and are therefore commonly researched for gas treatment or flow control purposes \cite{Brandenburg2017,Moreau2007,Mueller2007,Corke2010,HHKim2004}. SDBDs have the capability of being built in many unique geometrical configurations ranging in symmetry providing either a single axis or multiple axes for plasma propagation. They may also allow for either a single phase, anodic or cathodic plasma, or a dual phase ignition process.
    
    Throughout the 1990s SDBDs have been well investigated as potential actuators for gas flow control \cite{Brandenburg2017,HHKim2004,Moreau2007,Corke2010}. For such purposes an asymmetric geometry, where one electrode is offset from the opposite electrode and possibly completely submerged by the dielectric, is typically used \cite{Corke2010,Akishev2012,Audier2014,Biganzoli2012,Debien2012,GAO2017,Peng2019,Xiahua2016,Soloviev2017,Starikovskii2009,Unfer2010,Che2012,Hu2018,Shao2013,Soloviev2018,Opaits2008,Sato2019}. Much effort has been put into controlling the plasma behaviors, such as densities and surface charge deposition, and their corresponding aerodynamic effects from said SDBD configurations \cite{Opaits2008,Corke2010,Opaits2012,Audier2014,Sato2019}. It has also been shown that AC and pulsed waveforms can significantly modulate the plasma profiles (at positive and negative voltage phases) \cite{Akishev2012,Audier2014,Biganzoli2012,Che2012,Debien2012,Hu2018,Soloviev2017,Soloviev2018,Starikovskii2009,Unfer2010}. 
    
    In recent years, SDBDs have undergone extensive investigation for gas purification for industrial and environmental protection applications \cite{Brandenburg2017, Mueller2007,HHKim2004}. Absolutely calibrated two wavelength emission spectroscopy has been used in order to characterize a symmetric SDBD under tailored voltage waveforms \cite{Offerhaus2017,Offerhaus2018,Offerhaus2019}. The waveform under experimental investigation is a damped sine wave with multiple $\mu$s period, adjustable peak to peak voltage, and pulsed in the kHz regime. Additional emission spectroscopy, absorption spectroscopy, and Fourier transform infrared (FTIR) spectroscopy methods have also been used to measure various species densities and chemical modifications of cystine. Furthermore, flame ionization detectors, gas chromatography-mass spectroscopy, and ion energy analyzer quadrupole mass spectroscopy are all being used to investigate and characterize the conversion of volatile organic compounds into non-harmful and non-toxic compounds \cite{Schuecke2020}. Furthermore, the inclusion of pre gas heating and catalyst coatings are being investigated for higher conversion efficiencies \cite{Schuecke2020,Peters2021}.
    
    In many applications, like chemical processing and gas purification, the interaction between a plasma and a catalyst yields synergistic effects resulting in enhanced performances \cite{HHKim2004,HHKim1999}. As such, various structures of catalytic material are often inserted into traditional DBD reactors including, but not limited to: spheres, honeycombs, 3D fibre deposition structures and coatings of the dielectric barrier itself \cite{Zhang2018,HHKim1999}. The synergistic effect is obtained via two primary methods. Firstly, the altered geometry along with tailored voltage waveforms influence the discharge characteristics \cite{Brandenburg2017,HHKim2004,Zhang2018,HHKim2016,Zhang2015}. Secondly, the plasma distribution determines the effective contact area of the catalyst thereby altering the morphology and work function of the catalyst \cite{Neyts2014,Zhang2017}. This leads to a great importance on generating a controllable plasma density and spatial distribution \cite{Brandenburg2017,HHKim2004,Zhang2018,HHKim2016,Shang2019}.
    
    The above studies, although very interesting, were mostly based on experiments of submerged SDBDs where the plasma discharge is confined to one side of the dielectric plate providing investigations only into a single phase ignition process \cite{Akishev2012,Audier2014,Biganzoli2012,Corke2010,Debien2012,GAO2017,Moreau2007,Opaits2012,Peng2019,Xiahua2016,Shang2019,Shang2019,Soloviev2017,Starikovskii2009}. That is to say that only either an anodic or cathodic phase plasma is present, but never both simultaneously. This single phase nature limits the effective volume and surface area of the plasma which defines the effective catalytic surface area exposed to the plasma species in plasma enhanced catalysis. As such, the catalyst performance is potentially limited to a great extent in a single phase SDBD. In gas treatment conditions, an SDBD electrode system is very likely to be placed along the central plane parallel to gas flow in order to minimize flow resistance and increase the treatment volume. Under these conditions, it is very clear that utilizing an SDBD electrode system which ignites on both sides of the dielectric plate will improve the treatment volume, and as such efficiency of the process.
    
    Unfortunately, most theoretical investigations utilizing circuit models \cite{Pipa2012,Peeters2014,Pipa2020_PowerDBDEQC}, global models, molecular dynamic models \cite{Neyts2014}, fluid models \cite{Che2012,Peng2019,Soloviev2018}, and even particle-in-cell/Monte Carlo collision (PIC/MCC) models \cite{Zhang2015,Zhang2017,Zhang2018} of (S)DBDs and packed bed reactors provide limited insights into the underlying mechanisms of the plasma propagation \cite{Mujahid2018,Mujahid2020,mujahid2020Propagation}. No contributions on the theoretical investigation of a dual phase symmetric SDBD could be found by the authors, pointing to a significant lacking of knowledge of such configurations is present. The inherent mechanisms behind the evolution of the plasma discharge in asymmetric and even more so symmetric SDBDs is still not fully understood. It is not yet clear how a simultaneous positive and negative surface streamer (above and below the dielectric) can interact with each other, and to what extent, if any, do they enhance one another. It is not clear how the streamers respond to tailored voltage waveforms, nor what the optimized conditions are for generating large treatment volumes. It is unknown to what extent the surface streamers interact with an active surface such as a catalyst. These are crucial pieces of information to ensure good plasma enhanced catalysis performance. Additionally, many experiments, such as optical emission spectroscopy, still have open questions as to whether the results are more representative of the streamer bulk or the highly dynamic streamer head. These concerns demand a more detailed simulation for the dynamic behavior of the positive and negative streamers in a dual phase symmetric SDBD during the ignition process.
    
    \begin{figure}[t]
        \centering
        \includegraphics[width=0.4425\textwidth]{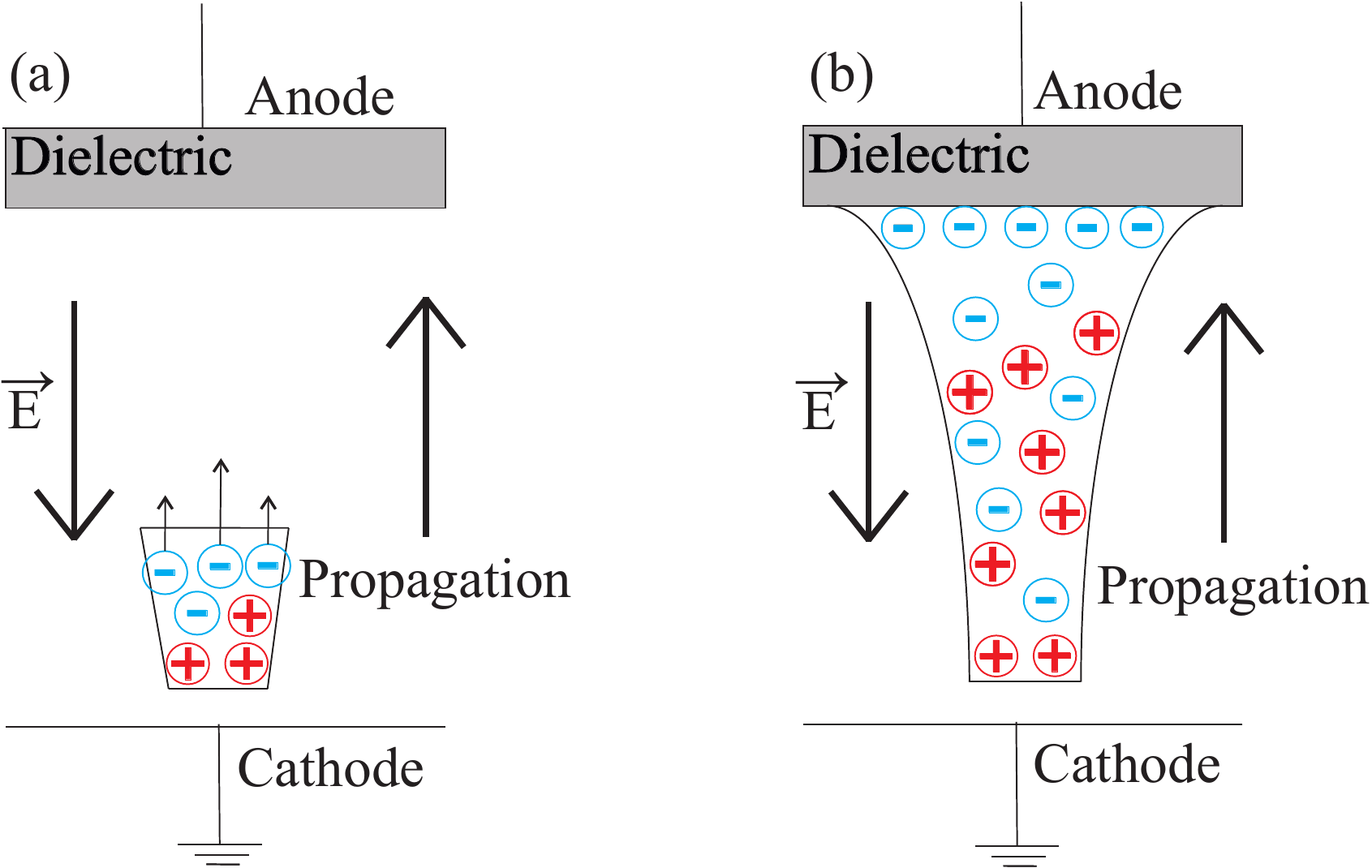}
        \caption{Schematic detailing the negative streamer formed via an anode oriented electron avalanche.}
        \label{fig:NegativeStreamer}
    \end{figure}
    
    Therefore, in the present work we computationally investigate the plasma propagation of a symmetric, dual phase SDBD, hereby referred to as the twin SDBD, under various voltage waveform conditions. The particular geometry of the twin SDBD ensures that both an anodic and cathodic phase plasma are simultaneously ignited, separated by the dielectric barrier, and are physically symmetric about the metallic electrodes. The symmetric geometry does not only give rise to a higher plasma surface coverage, but also enables a direct comparison between the positive streamers on the anode side versus the negative streamers on the cathode side as well as the interaction between the two. The numerical investigations are carried out by means of a 2D PIC/MCC simulation software known as VSim, a multi-physics simulation tool, which combines the Finite-Difference Time-Domain (FDTD), PIC, and Charged Fluid (Finite Volume) methods for simulating electrical gas discharges. \cite{NIETER2004}. The insights provided by this work are not only applicable to the twin SDBD and similar geometries, but also to other SDBD geometries, asymmetric ones included via a deeper understanding of the streamer propagation and form.
    
    \begin{figure}[t]
        \centering
        \includegraphics[width=0.4425\textwidth]{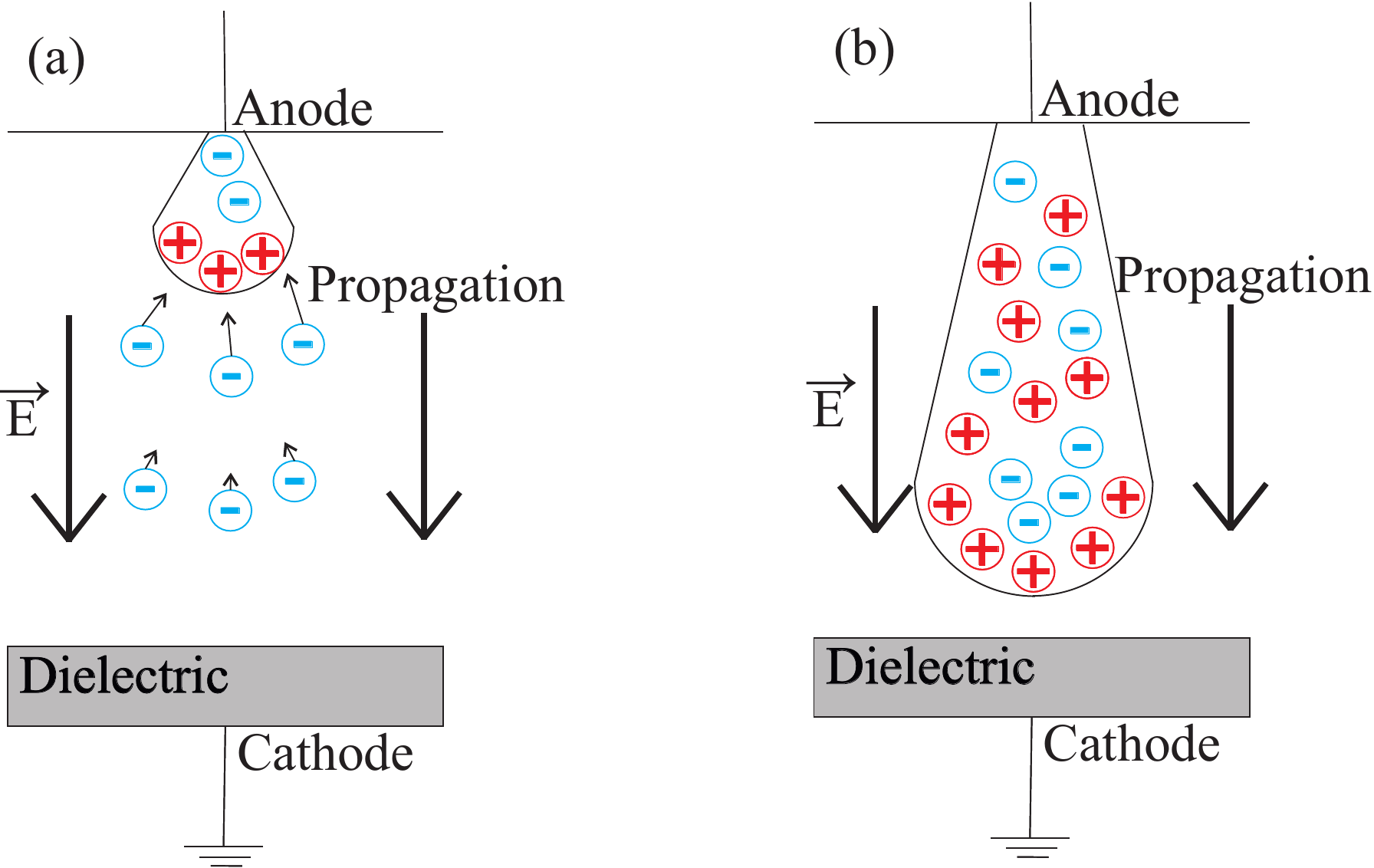}
        \caption{Schematic detailing the positive streamer, which forms via a cathode oriented propagation front.}
        \label{fig:PositiveStreamer}
    \end{figure}

    To provide a basis of understanding the streamer dynamics in a twin SDBD, that will be revealed in this work, we briefly recall the fundamentals of positive and negative streamer dynamics in a DBD. A negative streamer, see \cref{fig:NegativeStreamer}, ignites through an anode oriented electron avalanche: electrons, which are accelerated against the direction of the electric field, collide with the background gas. Ionization takes place causing an exponential growth of electrons and ions, creating a quasineutral bulk plasma that propagates from the cathode to the anode. A positive streamer, see \cref{fig:PositiveStreamer}, is also created via electron collisions, but is somewhat more complex. The cathode oriented positively charged streamer head attracts the electrons which cause ionization in front of the streamer head, resulting in an ionization wave. This ionization wave propagates from the anode to the cathode, leaving behind a quasineutral bulk plasma. Branches may form from the streamer head creating additional ionization waves; branching is more readily observed in gas mixtures that are susceptible to self induced photo ionization. Under short timescales, a few nanoseconds and less, a feature very similar to a low pressure sheath forms. The positive streamer head floats above the cathode due to an absence of available electrons, thus creating a region with a very strong electric field. Given an appropriate amount of time, the positive ions do reach the cathode due to their own velocities. At the dielectric(s), any charges that reach the surface adhere to it and charge it. These surface charges repel incoming like charges along the surface, causing both positive and negative streamers to spread out. Due to the lightweight electrons, this effect is more prominent in negative streamers; however, the floating nature of positive streamers can also facilitate a similar effect. For a deeper understanding we refer the reader to Nijdam \textit{et. al.} and to Zhang \textit{et.al.} \cite{Nijdam2020,Zhang2021} where the dynamics of positive and negative streamers of a VDBD via PIC/MCC simulations are detailed.

    \begin{figure}[t]
        \centering
        \includegraphics[width=0.4425\textwidth]{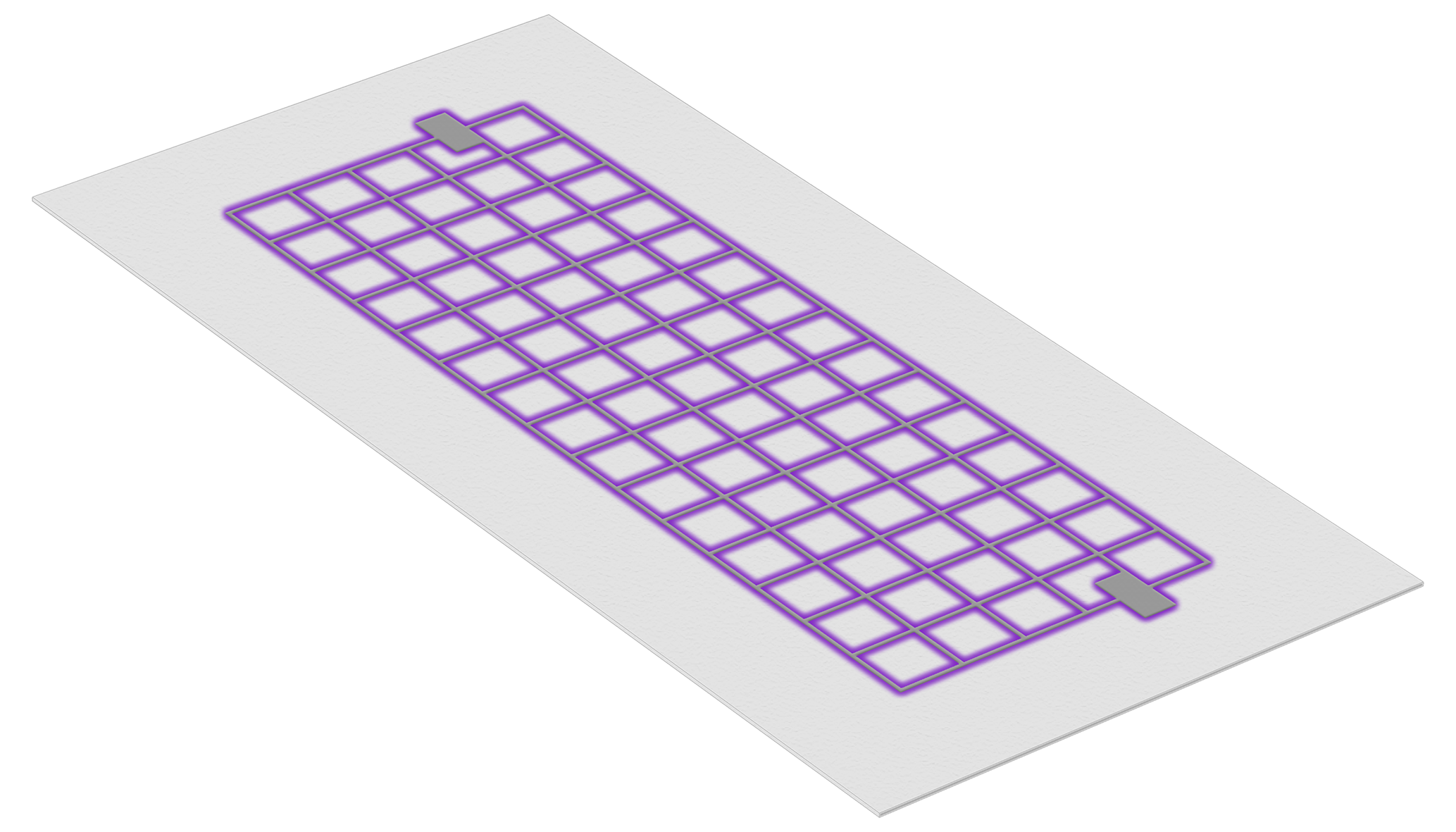}
        \caption{Computer generated graphic showing the physical structure of the SDBD electrode under consideration. A metallic lattice (dark grey structure) is printed symmetrically on both the top (visible) and bottom (hidden) faces of the Al$_2$O$_3$ dielectric barrier (light grey material). Due to the strong curvature of the electric field lines when under operation, the plasma (purple structure) ignites along the edges of the metallic lattice.}
        \label{fig:Electrode}
    \end{figure}

    This paper is structured as follows: First in \cref{Model} the computational model and geometry are described. Following this, in \cref{Results} the results of the various simulations are presented: the DC results in sub-\cref{SingleStreamers,DualStreamers}, and the AC results in sub-\cref{ACStreamers}. Finally, in \cref{Conclusion} our closing remarks and conclusions are discussed.

\section{Computational model}
\label{Model}
     \begin{figure}[t]
        \centering
        \includegraphics[width=0.4425\textwidth]{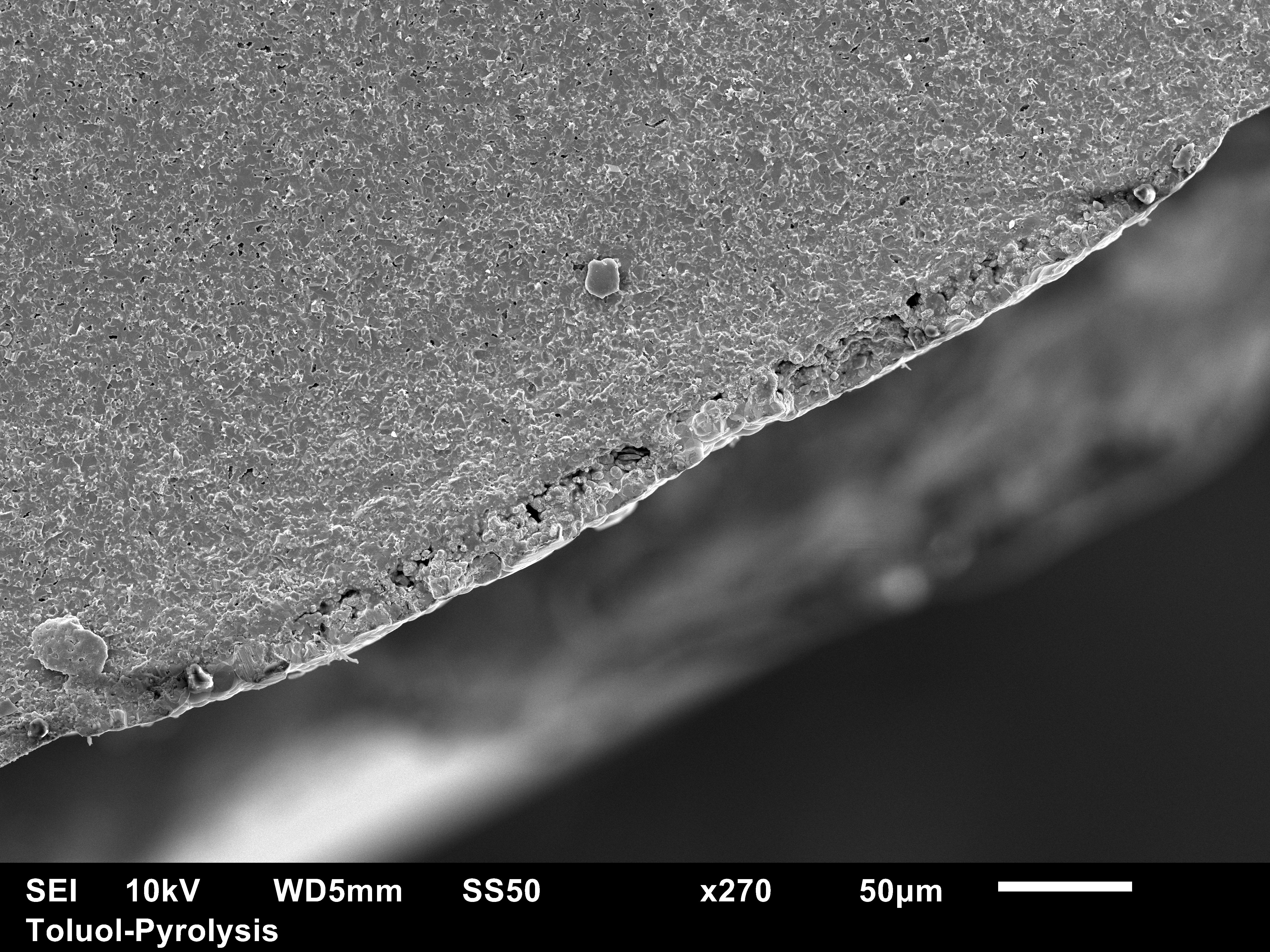}
        \caption{SEM image of electrode cross section. The bulk, homologous material is the Al$_2$O$_3$ dielectric. The hump like structure with larger grains is the metallic electrode trace.}
        \label{fig:SEM}
    \end{figure}

    The geometry to be simulated is chosen to resemble that of the twin SDBD electrode intended for use in gas treatment applications and was first experimentally presented in \cite{Offerhaus2017} and subsequently in \cite{Offerhaus2018,Offerhaus2019,Kogelheide2019,Schuecke2020}. The authors defer the readers to these references for a detailed description of the twin SDBD system under question. It is important to reiterate that this device consists of a dielectric plate, with metallic grids placed on the surface of the dielectric on both sides. A computer rendered sketch of the system can be seen in \cref{fig:Electrode}. These grids serve as electrodes. The system is built with both a geometric and electrical symmetry, such that both a positive and negative streamer are simultaneously ignited on either side of the dielectric under any given sufficiently high voltage conditions, which thereby warrants the name ”twin SDBD”. The metallic traces of the electrode system have been imaged with a scanning electron microscope for a more accurate depiction of the electrodes within the simulations. An example image of the cross sectional view of the metallic traces can be seen in \cref{fig:SEM}, which shows the curved nature of the metallic traces located on the dielectric, which is included in the simulation.

    \subsection{Particle in Cell/Monte Carlo Collision model}
    \label{Sim Model}
        \begin{figure}[t]
            \centering
            \includegraphics[width=0.4425\textwidth]{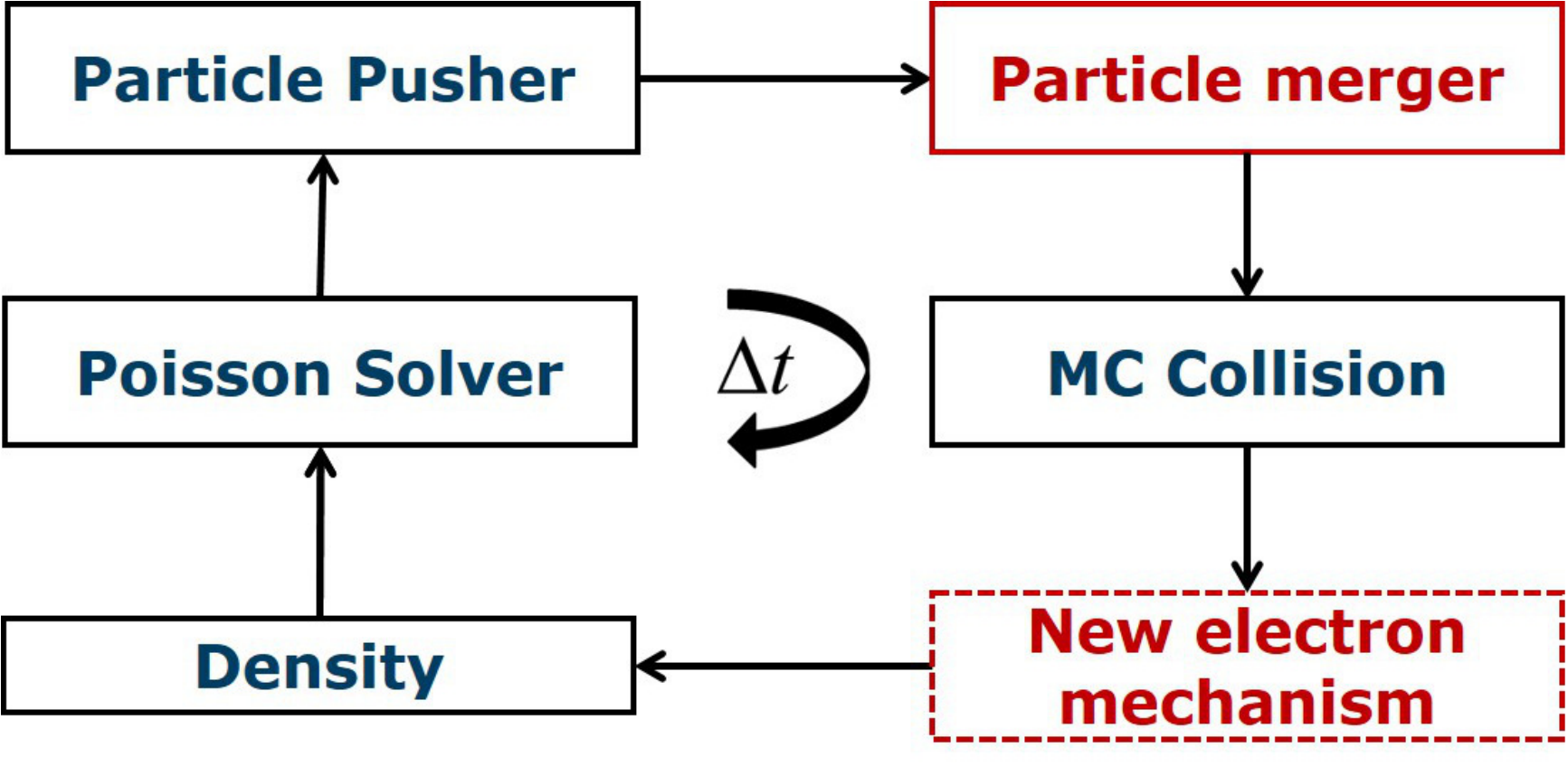}
            \caption{Logic flow diagram of the PIC/MCC algorithim. One complete loop of the flow diagram represents one time stamp of the PIC/MCC code. During each successive change in the time step of the simulation, all sub algorithms are performed. Particles are pushed, merged, collided, generated, the densities are determined, and analyzed for electrical forces.}
            \label{fig:ModelFlow}
        \end{figure}
        
        A 2D PIC/MCC model is used to study the plasma propagation of the twin SDBD based on the VSim simulation software \cite{NIETER2004}. VSim is being widely used and has been validated \cite{NIETER2004,Zhang2015,Zhang2017}. As these investigations taken place under similar conditions presented here (atmospheric pressure DBDs, nanosecond timescales and micrometer length scales), we operate under the assumption that our model is also valid. Additionally, the usage of PIC/MCC simulations to investigate the COST-Jet at atmospheric pressure yield realistic results that agree well with experiments, \cite{Bischoff2018,Korolov2019,Korolov2020}, proving that PIC/MCC models can indeed be used at atmospheric conditions. The PIC/MCC simulations performed in VSim are based on an explicit solver and the electrostatic approximation of Maxwell's equations, which were described in detail in \cite{Birdsall1991}. The PIC/MCC model takes advantage of accounting for the detailed kinetic behavior of charged particles which may be important for the evolution of electron avalanches and branching mechanisms, and therefore, the plasma streamer profiles. Air at atmospheric pressure is used as the discharge gas, with a constant density of background molecules, $80\,\%$ N$_2$ and $20\,\%$ O$_2$, at $300\,$K. Free electrons, N$_2^+$, O$_2^+$ and O$_2^-$ ions are traced throughout the simulation, which are represented as super-particles, i.e. one super-particle corresponds to a certain number of real particles defined by their numerical weighting, initially starting at $20\cdot10^3$ real particles per super particle \cite{Birdsall1991}.
        
        In order to numerically initiate the plasma discharge, a uniform distribution of seed electrons is placed within the free space of the simulated geometry. These seed electron super-particles have a density corresponding to $1\cdot10^{15}\,$m$^{-3}$. Realistically, seed electrons are present due to cosmic radiation and environmental photo-ionization producing background electrons, as well as remaining charges from previous plasma discharges. The initial electron density was chosen as such in order to increase the initial weighting of the super particles, and thereby the simulation speed. The high initial density increases the speed of the initial electron avalanches and streamer breakdown. As seen later on, maximum achieved densities are on the order of $1\cdot10^{22}\,$m$^{-3}$, which is much higher than the initial density; therefore, the final profiles and mechanisms would not change if a lower initial density was chosen. Thus, the high initial density serves to increasing the simulation speed while not altering the results of the simulations. It should be noted that the usage of uniform seed electrons does not consider local effects of previous discharges.
        
        As the plasma streamers evolve, the particle number of each considered species will rapidly increase due to the ionization avalanches. To account for this and to reduce the computation time, the weight of each super-particle is adaptive. A merger algorithm conserving both momentum and energy will combine same species super-particles when the number of said super-particles exceeds a threshold value of 10 super-particles respective to each cell of the simulation mesh. As the particle numbers only increase within the considered simulated time, no de-merger algorithm is implemented. This adaptive weight and merger algorithm is described in more detail in \cite{Zhang2017}.
        
         Elastic, excitation, ionization, and attachment collisions of electrons with O$_2$ and N$_2$ gas molecules make up the considered reaction mechanisms as explained in more detail by \cite{Zhang2017}. The corresponding cross sections and threshold energies are adopted from the LXCat database and literature \cite{LiebermannAndLichtenberg,Furman2002,A_V_Phelps1999,PANCHESHNYI2012,LXCATdatabase}. At the surface of the dielectric barrier, only electron absorption is considered, \textit{i.e.} no electron reflection or surface electron emission is considered. Reported in \cite{Zhang2015,Zhang2017}, the inclusion of secondary electron emission, SEE, surface coefficients do not significantly alter the form of the simulated positive streamers, due to the floating nature of the streamer head. The negative streamer; however, propagates along the surface of the dielectric barrier, and as such, SEE coefficients would be more critical. The inclusion of SEE coefficients would theoretically increase the number of "background" electrons available for streamer propagation, and as such the streamers would propagate faster; however, their forms should not strongly change. Additionally, due to the lower electric fields of the negative streamer and the very short considered timescales, the effect of ion induced SEE would be very limited within this investigation.
        
        \begin{figure*}[t]
            \centering
            \subfloat{
                \label{fig:Geom(a)}
                \includegraphics[width=0.885\textwidth]{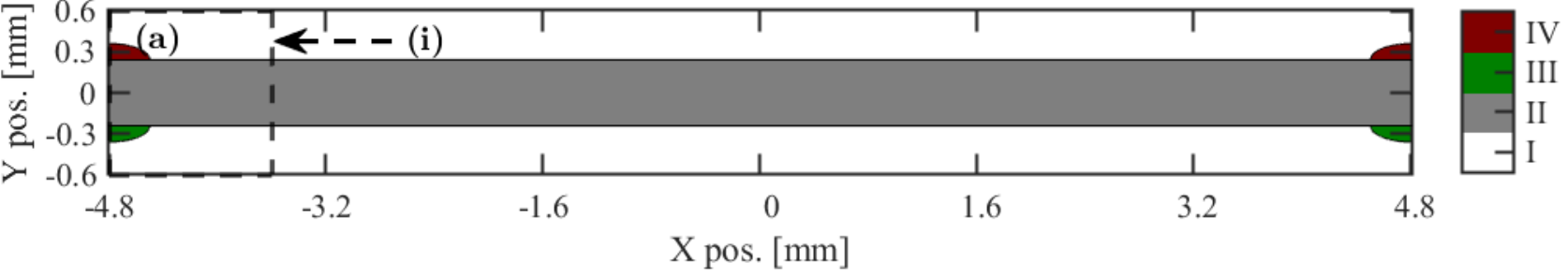}}
            \\
            \subfloat{
                \label{fig:Geom(b)}
                \includegraphics[width=0.885\textwidth]{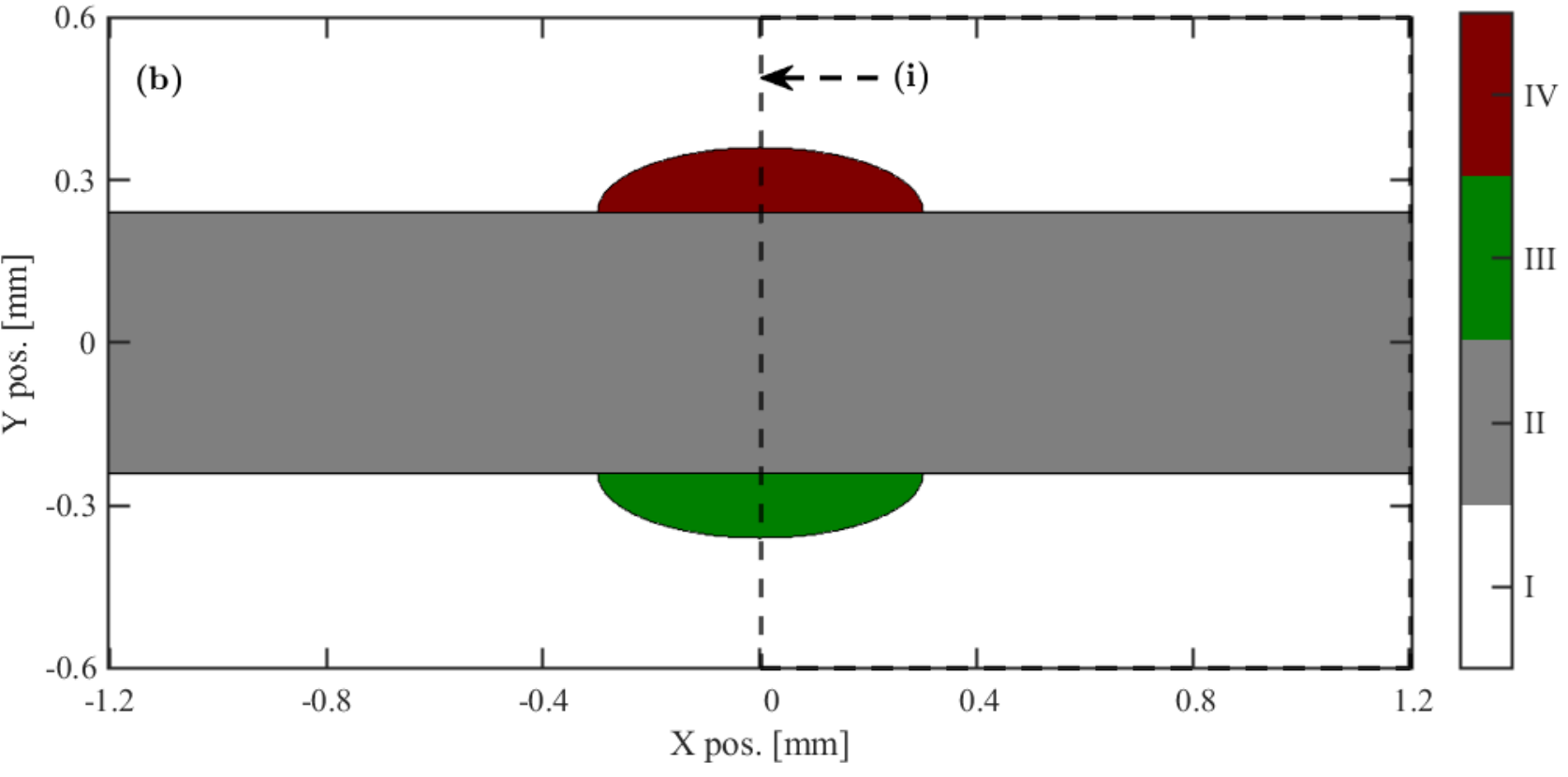}}
            \caption{Schematic of the simulation regimes. Subfigure (a) and (b) correspond to the DC and AC simulated geometries respectively. The color scale corresponds to the different materials as follows: I) air ($80\,\%$ N$_2$ and $20\,\%$ O$_2$), II) Al$_2$O$_3$ dielectric, III) grounded electrode, IV) powered electrode. The boxed in regions denoted with (i) correspond to the regions that are presented in greater detail for the rest of the publication.}
            \label{fig:Geometry}
        \end{figure*}
        
         With each successive timestamp of the model, a particle pusher, particle merger, and Monte Carlo collision algorithms for all particle species follow in succession. After the collisions, a new electron super particle is added to the simulation regime, the density of each cell is calculated, and Poisson's equation is solved in order to get the electric forces being applied to each particle, after which the cycle repeats. A diagram of the general flow is shown in \cref{fig:ModelFlow}.
        
    \subsection{Simulated geometry}
    \label{Sim Geometry}
        
        The geometry to be simulated is a cross section of the twin SDBD described in \cite{Offerhaus2017,Offerhaus2018,Offerhaus2019,Kogelheide2019,Schuecke2020}, and shown in \cref{fig:Electrode}. The twin SDBD simultaneously produces positive and negative phased plasma streamers along the edges of the metallic traces; however, the two phases are separated by the Al$_2$O$_3$ dielectric barrier. On either side of the dielectric barrier, ignition on opposite edges of the respective metallic trace can be considered as two individual but same-phased streamers. Two different simulation geometries, referred to as geometry(a) and geometry(b), are considered in order to appropriately resolve the interaction of both the same-phased and respectively opposite-phased plasma streamers. Simulation geometry(a) and simulation geometry(b) are presented in \cref{fig:Geometry}. In total geometry(a) contains a 2D plane that is $9.6\,$mm x $1.2\,$mm in Cartesian X and Y coordinates. The plane is uniformly divided into square cells with unit length of $2.4\,\mu$m resulting in a square lattice of 4000 x 500 cells. The grid size was chosen based off of the Courant limit, $c\cdot dt<dx$, where $c$ is the speed of light and $dx$ is the grid size. Geometry(b) utilizes the same size grid cell, but uses only 1000 x 500 cells resulting in a total width of $2.4\,$mm. For ease of comparison, results from a zoomed in region of size 500 x 500 cells from both simulated geometries are presented for the rest of the paper. The respective regions are outlined by a dashed line and annotated with $(i)$ in \cref{fig:Geometry}.
        
        \begin{figure*}[t]
            \centering
            \includegraphics[width=0.885\textwidth]{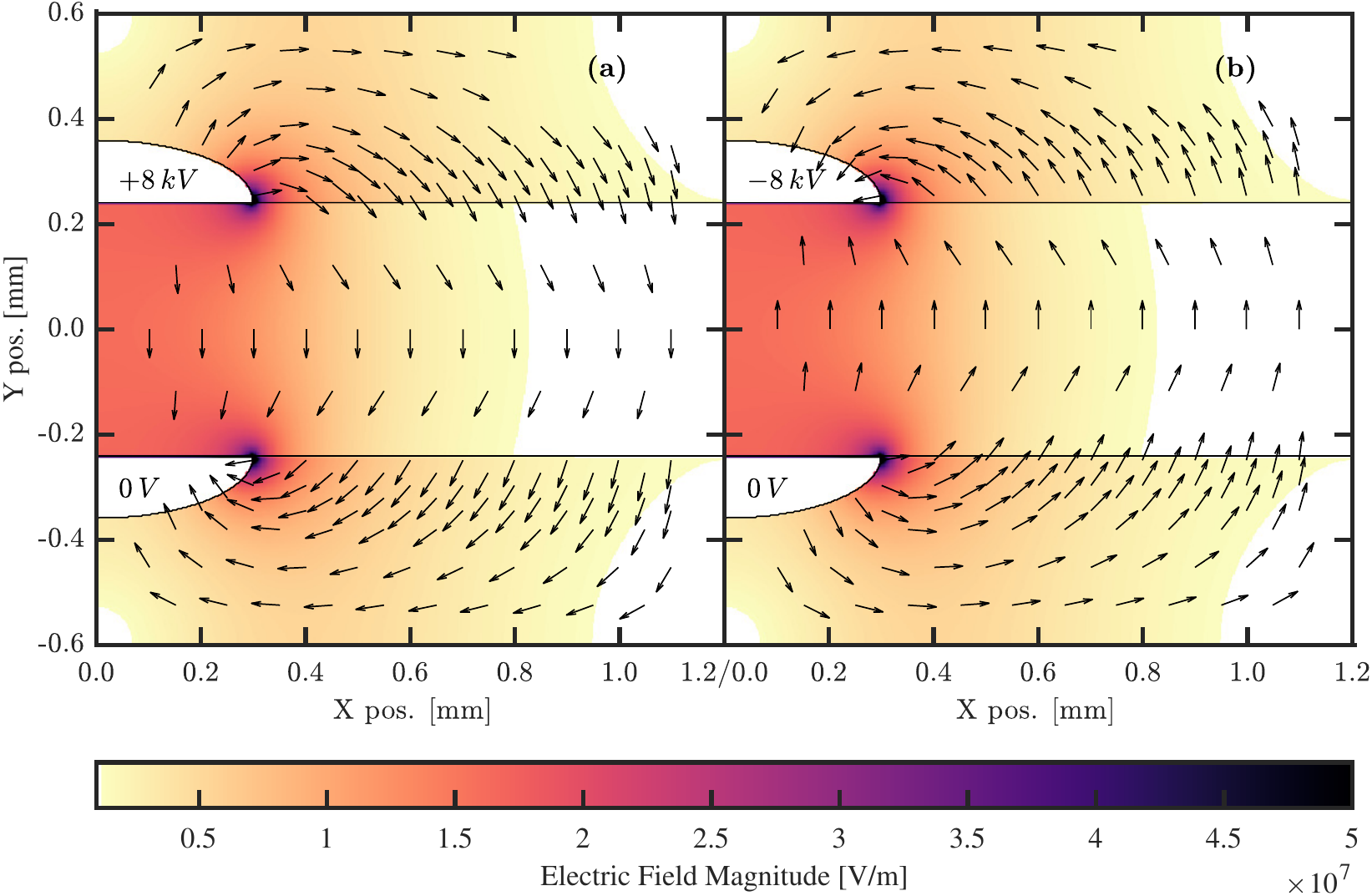}
            \caption{Electric field distribution of the simulated electrode geometries for an applied $+8\,$kV and $-8\,$kV potential in (a) and (b) respectively. The magnitude of the electric field is plotted on a linear intensity color scale, where the threshold value for the minimum intensity is chosen to be $1\cdot10^{6}\,$V/m. The normalized direction of the electric field is shown via the vector field.}
            \label{fig:EField}
            
            \includegraphics[width=0.885\textwidth]{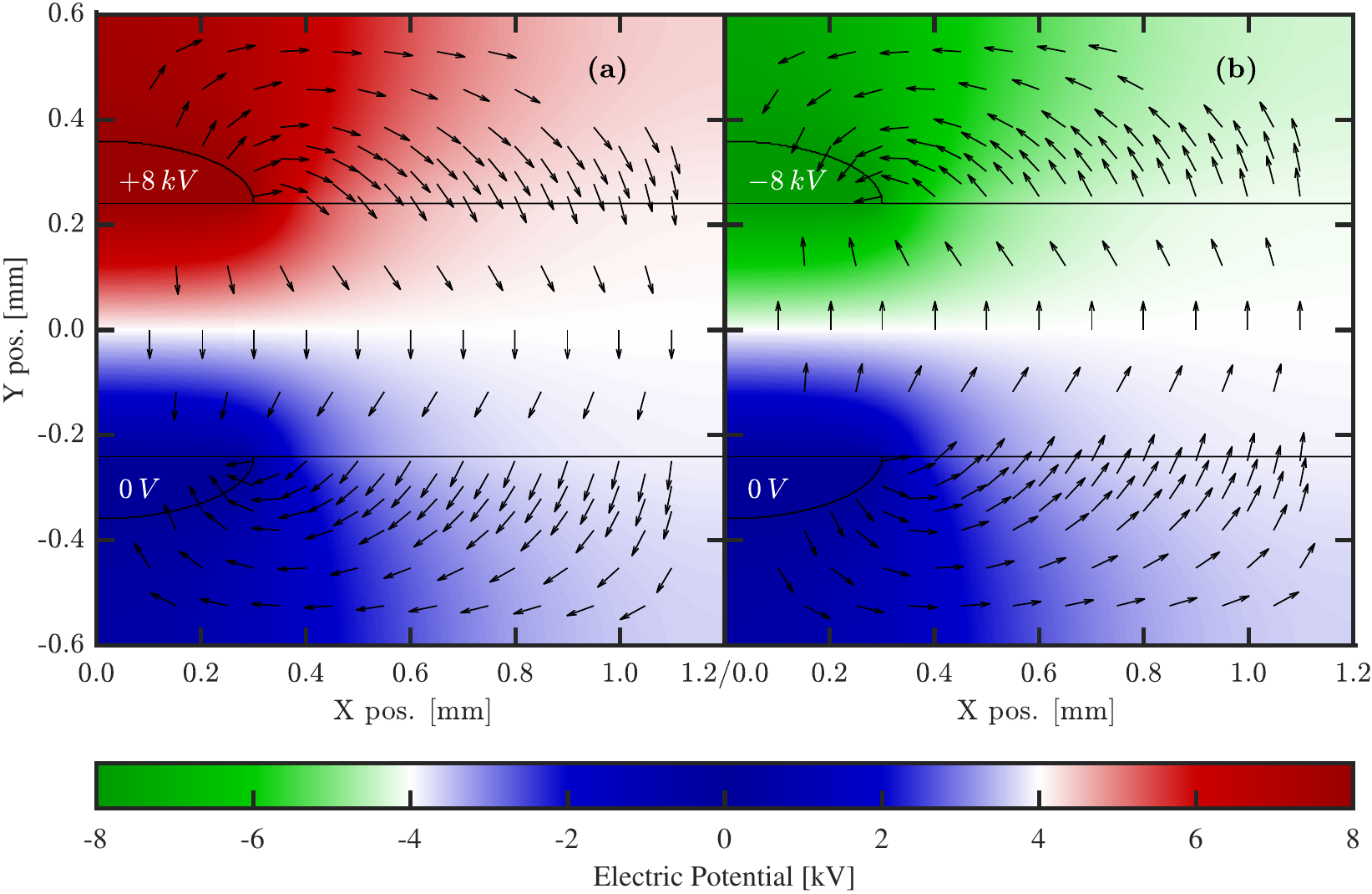}
            \caption{Electric potential distribution of the simulated electrode geometries for an applied $+8\,$kV and $-8\,$kV potential in (a) and (b) respectively. The electric potential is plotted on a linear intensity color scale. Additionally, the normalized direction of the electric field is shown via the vector field.}
            \label{fig:EPotential}
        \end{figure*}
        
        Firstly, to investigate the interactivity of two same-phase streamers, positive-positive or negative-negative, two anodes and two cathodes are included in simulation geometry(a). The two same-phase electrodes are simulated with the same potential under DC conditions and are separated in the X-direction by $9.5\,$mm, corresponding to the distance between the edges of two neighboring and parallel metallic traces of the physical electrode. In order to minimize the computational time, the X boundaries of geometry(a) correspond to the vertical center lines of the metallic traces. Simulation geometry(a) may be seen in \cref{fig:Geom(a)}. Later, in \cref{SingleStreamers,DualStreamers} it is deduced that minimal interactivity is observed between two same-phase streamers. This is due to the limited spatial propagation of the plasma streamers on the considered timescales. Therefore, it is appropriate to simulate a section centered about just one metallic trace under the same timescales, thus a second simulation geometry is investigated. In simulation geometry(b) only one set of electrodes is considered, is only simulated under AC conditions, and is centered about the X-axis with the walls being $1.2\,$mm away from either side of their center line. Concerns about the reduced simulation domain having an effect on the calculated electric field strengths are mitigated by the naturally very fast reducing field strength as a function of the square of the distance from the electrodes. The usage of Neumann boundary conditions additionally improves the accuracy, as the simulation walls are not forced to a specific potential. Simulation geometry(b) may be seen in \cref{fig:Geom(b)}.
        
        Both considered geometries of the 2D PIC/MCC model represent a cross sectional view of the electrode structure, where the anodes and cathodes are separated along the Y-axis by the dielectric barrier. The dielectric is located in the middle of the Y-axis, was chosen to be $0.500\,$mm thick and expands the whole X-direction, and is simulated with a dielectric constant of 9. In this representation, the Z-direction would equate to the length (or width) of the physical electrode setup but is mathematically treated as constant/homogeneous. This results in a simulation regime that is most valid for a planar section in the middle of any grid structure. In both geometries, the electrode structure itself is a geometrical composition of multiple tangent arcs resulting in a "hump" like structure. This electrode structure is used to approximate the real geometric structure of the metallic traces which can be seen in \cref{fig:SEM}. It should be noted that the simulated aspect ratios of the electrode thickness and width to the dielectric thickness is significantly different from reality; however, this was chosen as such in order to avoid numerical issues which would arise from using an appropriately sized simulation grid for realistic aspect ratios. Furthermore, the reduced dielectric thickness of the simulations versus the actual electrode configuration should not lead to any major differences in the interpretations of this paper, as it is the surface of the dielectric that plays a much more important role. By using a reduced dielectric thickness, we are able to increase the number of computational cells available for the plasma propagation, without increasing the entire simulation domain.
        
        Particle densities and electric fields are resolved using a cutting-cell technique in order to handle the irregular geometry, through contributions of neighboring cells. The authors refer the reader to references \cite{Smithe2008,Meierbachtol2015,loverich2010} for more information. Neumann boundary conditions are used in all directions to ensure a smooth electric potential distribution at the boundaries of the simulation walls. The timesteps are non adaptive and fixed at $2\cdot10^{-13}\,$s. Similar to \cite{Likhanskii2010}, a singular new electron super-particle is randomly added to the simulation domain at each timestep in order to account for random events such as cosmic radiation, photo-ionization, \textit{etc.} as described in \cite{Ebert2006,E_M_van_Veldhuizen2002,Qiu2017}. These random events are beyond the scope of the available VSim functions. The seed electrons, both background and newly loaded electrons, are both sufficient in the simulation region to support streamer propagation as well as to not interfere with the plasma bulk as they are far fewer compared to the generated plasma. The generated plasma density profile is also much smaller than the simulation domain in both considered geometries.
        
    \subsection{Waveform variation}
    \label{Waveform}
        In all considered simulations and both geometries, the electrode(s) above the dielectric barrier are treated as the powered electrode(s) while the bottom electrode(s) are held constant at $0\,$V. This choice is arbitrary and due to the physical symmetry of the system would provide only mirrored results if the opposite choice, either inverse polarity and/or choice of powered electrode, was made. Initially, a constant positive $8\,$kV potential is applied to geometry(a), thus the two powered electrodes take the role of the anodes while the bottom two are the cathodes. The initial electric field distribution can be seen in \cref{fig:EField}(a) and the initial potential distribution can be seen in \cref{fig:EPotential}(a). Within both figures, the magnitudes of the presented quantity are shown via the color scale, and the normalized direction of the electric field are additionally presented for further clarity. The normalized direction is presented as a vector field, where the X and Y directions of the vectors are the normalized X and Y values of the electric field at that grid cell. Naturally, the magnitude of the electric field is obtained from the square root of the sum of the X and Y components squared: $E_{mag} = \sqrt{E_X^2 + E_Y^2}$.

        First, in order to investigate solely the role of the positive streamers, only the top half of the simulation area is seeded with the initial electrons. Likewise, the bottom half is subsequently seeded in a second simulation in order to solely investigate the negative streamers. Third, both halves are identically seeded thereby investigating the interplay and differences of both discharges igniting simultaneously under the DC voltage conditions. These three conditions are applied to geometry(a) only. Lastly, a varying voltage waveform is investigated.
        
        \begin{figure}[t]
            \centering
            \includegraphics[width=0.4425\textwidth]{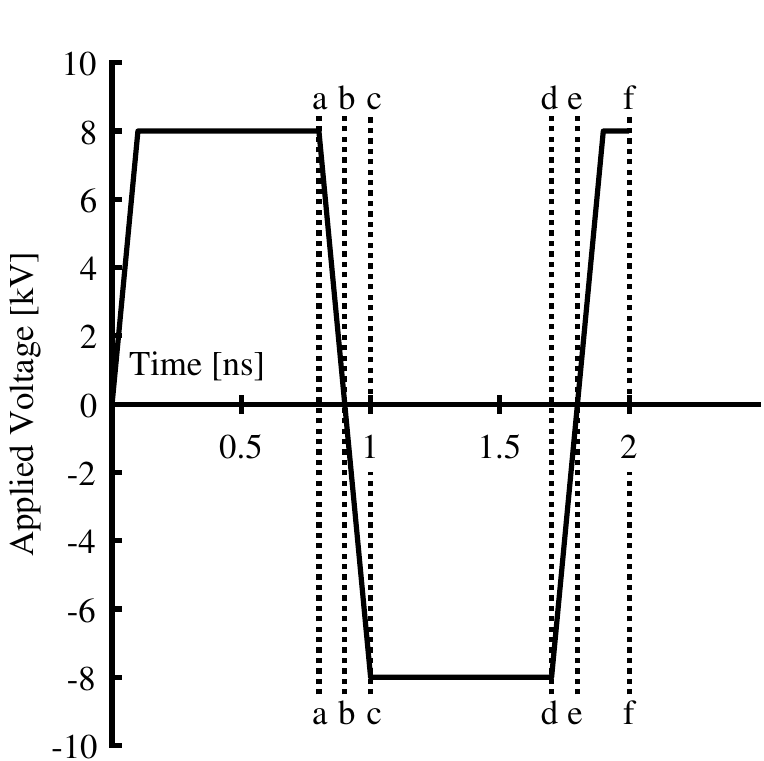}
            \caption{Applied voltage waveform of the AC simulations. Dashed lines labeled a through f at 0.8, 0.9, 1.0, 1.7, 1.8, and 2.0$\,$ns respectively represent the timestamps at which results are presented in \cref{DualStreamers}.}
            \label{fig:ACPulseform}
        \end{figure}

        Geometry(b) is only investigated under the AC conditions shown in \cref{fig:ACPulseform}. Under these conditions, the role of the anode and cathode switches twice; thereby giving insights into the extreme dynamics of fast voltage streamer switching. Initially, the applied voltage potential sharply rises within $0.1\,$ns to the $8\,$kV maximum which is then held constant for $0.7\,$ns. During this time, the anode is located on the top side of the dielectric barrier. At $0.8\,$ns, the voltage is decreased at the same rate, $80\,$kV/s, reaching the minimum applied voltage of $-8\,$kV at $1\,$ns making the top side of the dielectric barrier the cathode. Again, this minimum value is held constant for $0.7\,$ns until switching back to the positive $8\,$kV potential, again switching the location of the anode and cathode. Without considering any plasma propagation, the base electric field distribution for both a positive and negative applied potential are shown in \cref{fig:EField} and the equivalent potential distribution can be seen in \cref{fig:EPotential}.

        All conditions are simulated for up to a maximum of $2\,$ns, thereby only revealing the inception phase of the streamers. The insights revealed within the \cref{Results} are consistent with other PIC/MCC models investigating DBD streamers in structured and porous catalytic surfaces \cite{Zhang2018,Zhang2018Porous}, which also are simulated in the ns timescales. Additionally, the phenomenon of a floating positive surface discharge is also observed in various fluid models \cite{Babaeva2016,Yan2014}. Therefore, the authors believe the results presented throughout this paper, even given the short time scales, are reasonable. The results reported below are meant for a qualitative understanding of the streamer dynamics in a twin SDBD. The general conclusions for more natural voltage waveforms, such as continuous sine waves, can be drawn, and could warrant further studies considering a real RF source. However, the results obtained in this work are particularly relevant for tailored voltage waveforms, which is a hot topic of current research and is trending towards shorter pulses and steeper rise times.

\section{Results and Discussion}
\label{Results}
    \subsection{Single Streamer Dynamics} \label{SingleStreamers}
        \begin{figure*}[t]
            \centering
            \includegraphics[width=0.885\textwidth]{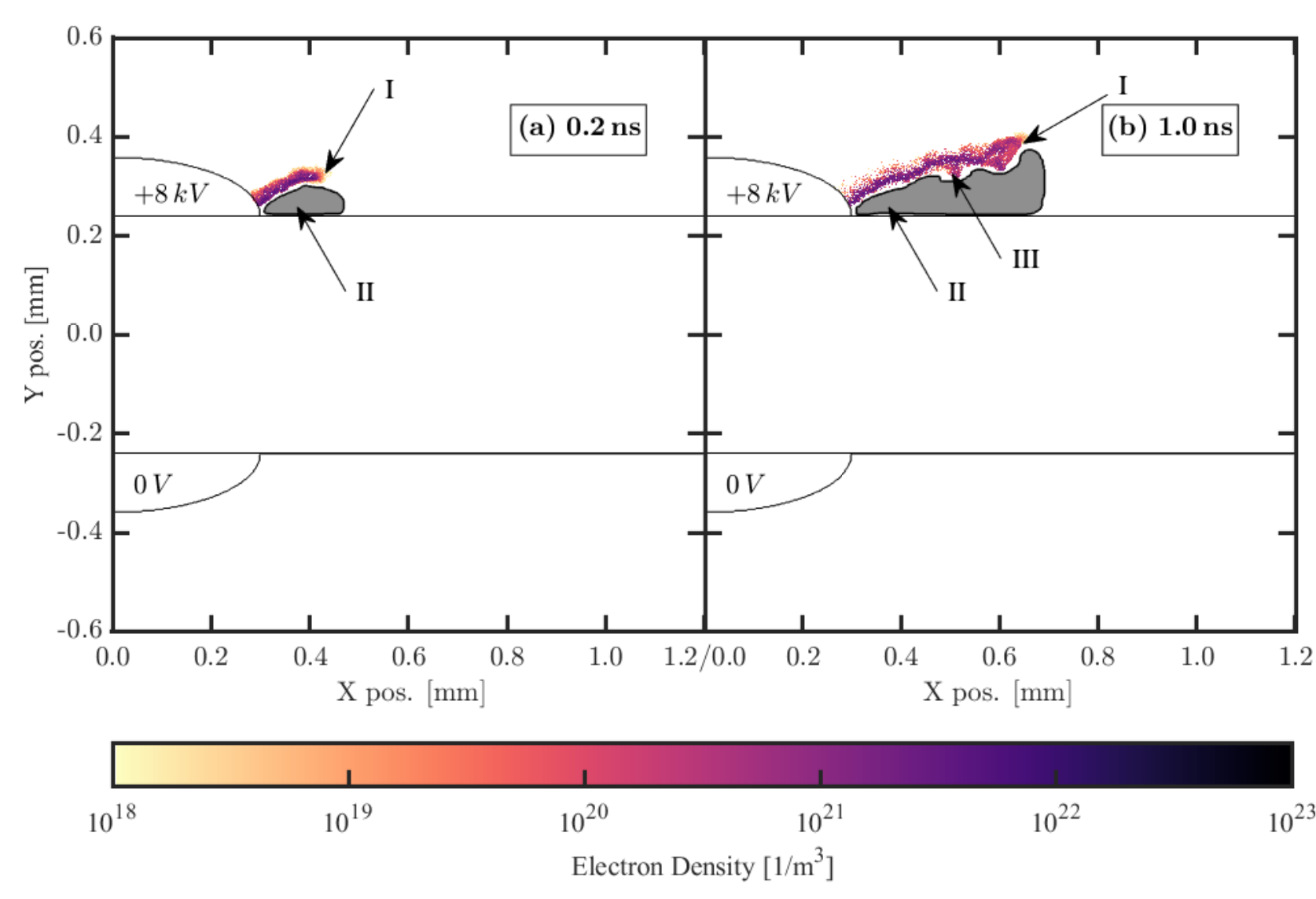}
            \caption{Spatial profiles of the electron density plotted on a logarithmic intensity scale of the positive streamer simulations with constant voltage. Sub figures (a) and (b) correspond to the timestamps of 0.2 and $1.0\,$ns, respectively. Features of importance are labeled, where the annotations are as follows: I) positively charged streamer head leading to streamer propagation, II) shaded region showing location of electron depletion, \textit{i.e.} sheath like feature, III) potential/failed positive streamer branch.}
            \label{fig:DC PositiveStreamer}
        \end{figure*}
        
        Under the 8kV DC conditions with seed electrons present only on the anodic side of geometry(a), the propagation of an anodic phased plasma streamer, also known as positive streamer is simulated and presented in \cref{fig:DC PositiveStreamer}. The initial electric field distribution is shown in \cref{fig:EField}(a) and the initial electric potential distribution is shown in \cref{fig:EPotential}(a). Under these conditions, a cathode oriented positively charged streamer head that is able to freely move from the metallic anode to the dielectric surface is able to form.
        
        \begin{figure*}[t]
            \centering
            \includegraphics[width=0.885\textwidth]{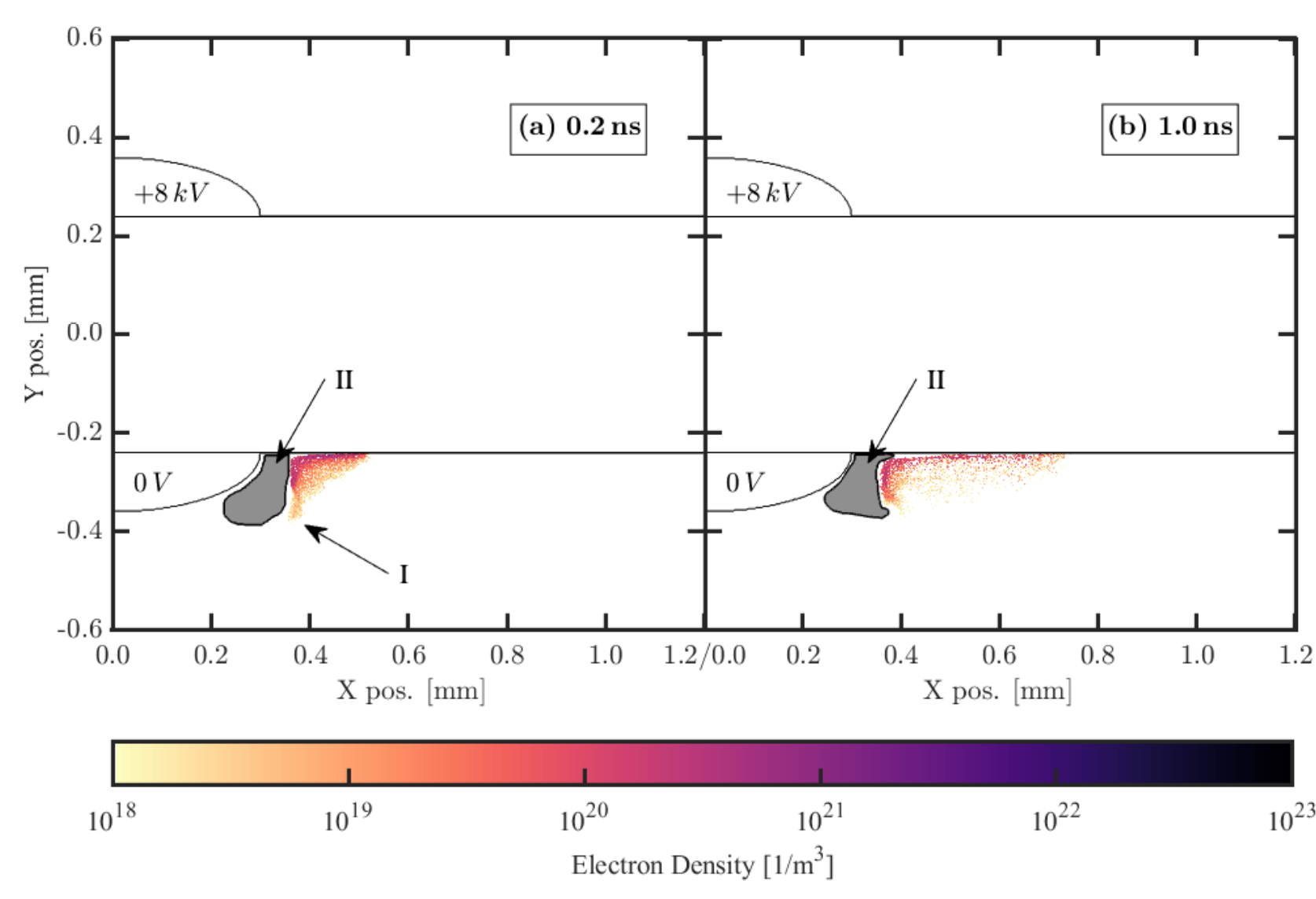}
            \caption{Spatial profiles of the electron density plotted on a logarithmic intensity scale of the negative streamer simulations with constant voltage. Sub figures (a) and (b) correspond to the timestamps of 0.2 and $1.0\,$ns, respectively. Features of importance are labeled, where the annotations are as follows: I) positively charged region leading to positive streamer like propagation, II) shaded region showing location of electron depletion, \textit{i.e.} sheath like feature.}
            \label{fig:DC NegativeStreamer}
        \end{figure*}
        
        The streamer structure is anchored to the anode just above where the highest electric fields are located. It would be expected that the anchoring would take place at the location of the highest electric field; however, under these conditions this is located at the intersection of the electrode and the dielectric surface. At this point, and immediately next to it, due to the strong curvature of the electric field, electrons do not have enough space to gain sufficient energy for ionization. Multiple executions of the simulation produce anchored positions at the same location; furthermore, the anchor position is also at a symmetrical position on the opposite anode, which is not presented in \cref{fig:DC PositiveStreamer}. This suggests that the anchor is positioning itself based on the strong curvature of the anode, and not through the randomness of the ionization events. Indeed, when looking at the curvature of the simulated electrode, it appears as if the plasma is next to the strongest curvature. Under no conditions did the simulated positive streamers extend a significant amount into the X-direction, such that interactions between the two positive streamers do not need to be considered.
        
        At $0.2\,$ns the positive streamer has advanced $0.12\,$mm meaning a propagation speed of $0.62\,$mm/ns. By the end of the simulated time, $1.0\,$ns, the streamer had stopped propagating a significant amount. The positive streamer had reached a propagation distance of $0.31\,$mm resulting in an averaged speed of $0.31\,$mm/ns. The actual instantaneous speed of the streamer would be significantly slower at this timestamp, as the average includes the faster propagation of the early streamer. It was observed via multiple test executions that these propagation speeds and distances were highly dependent on the initial background electron density. With lower initial densities, the simulated streamer propagates a shorter distance. Likewise, larger background densities would result in faster speeds and longer propagation distances.
        
        Initially the positive streamer began to propagate along the electric field lines at an angle offset from the surface of the dielectric barrier. The positive streamer head, which is not directly visible in \cref{fig:DC PositiveStreamer}, forms in front of the streamer and along the bottom side between the bulk plasma and the dielectric barrier. The streamer head is annotated in \cref{fig:DC PositiveStreamer} with an arrow labeled (I). Between the dielectric barrier and the positively charged streamer head is located a sheath like region, annotated via (II), where free electrons are attracted to the streamer head; however, they do not have enough space in order to promote further propagation towards the dielectric. Therefore, the only direction possible is outwards along the X- and positive Y-directions, towards the center of the simulated area. As the streamer continues to propagate along this direction, the electric field gets weaker proportional to $1/r$ (in 2D) or $1/r^2$ (in 3D), where $r$ is the distance from the electrode. Thus the positive streamer is able to advance in a somewhat straight line, parallel to the initial trajectory, which is at some angle to the dielectric surface; under these presented conditions this trajectory angle was determined to be $20.6^\circ$. The further the streamer propagates, the more space is available for propagation into the negative Y-direction, towards the dielectric surface. Therefore, in \cref{fig:DC PositiveStreamer}(b), a potential branch had began to take shape, annotated with (III); however it is not able to fully develop. As the cathode is located underneath the positive streamer, that is the only location of the streamer head; therefore, no branching occurs above the streamer bulk.

            Due to the location of the failed branch in \cref{fig:DC PositiveStreamer}(b)(III), it would be extremely difficult to experimentally observe, and is noticeable within these simulations because of the kinetic nature of PIC/MCC models. Naturally, without experimental evidence, the reader might question the reality of whether branching forms or not at these orientations. The authors believe that the simulations are indeed accurate in predicting these features.

        In \cref{fig:DC NegativeStreamer} the same simulation conditions are presented, except the initial seed electrons are on the cathode side of the dielectric barrier, thus the negative streamer is simulated. The seed electrons are still accelerated in the opposite direction of the electric field lines shown in \cref{fig:EField}(a). An electron avalanche directed towards the anode initiates the discharge. Under these conditions the electrons are pushed towards the dielectric, where they begin to collect on and charge the surface of the dielectric. A positively charged spatial region forms next to the cathode, but is unable to anchor to the cathode, as it must float at some distance away from the cathode.
        
        \begin{figure*}[t]
            \centering
            \includegraphics[width=0.885\textwidth]{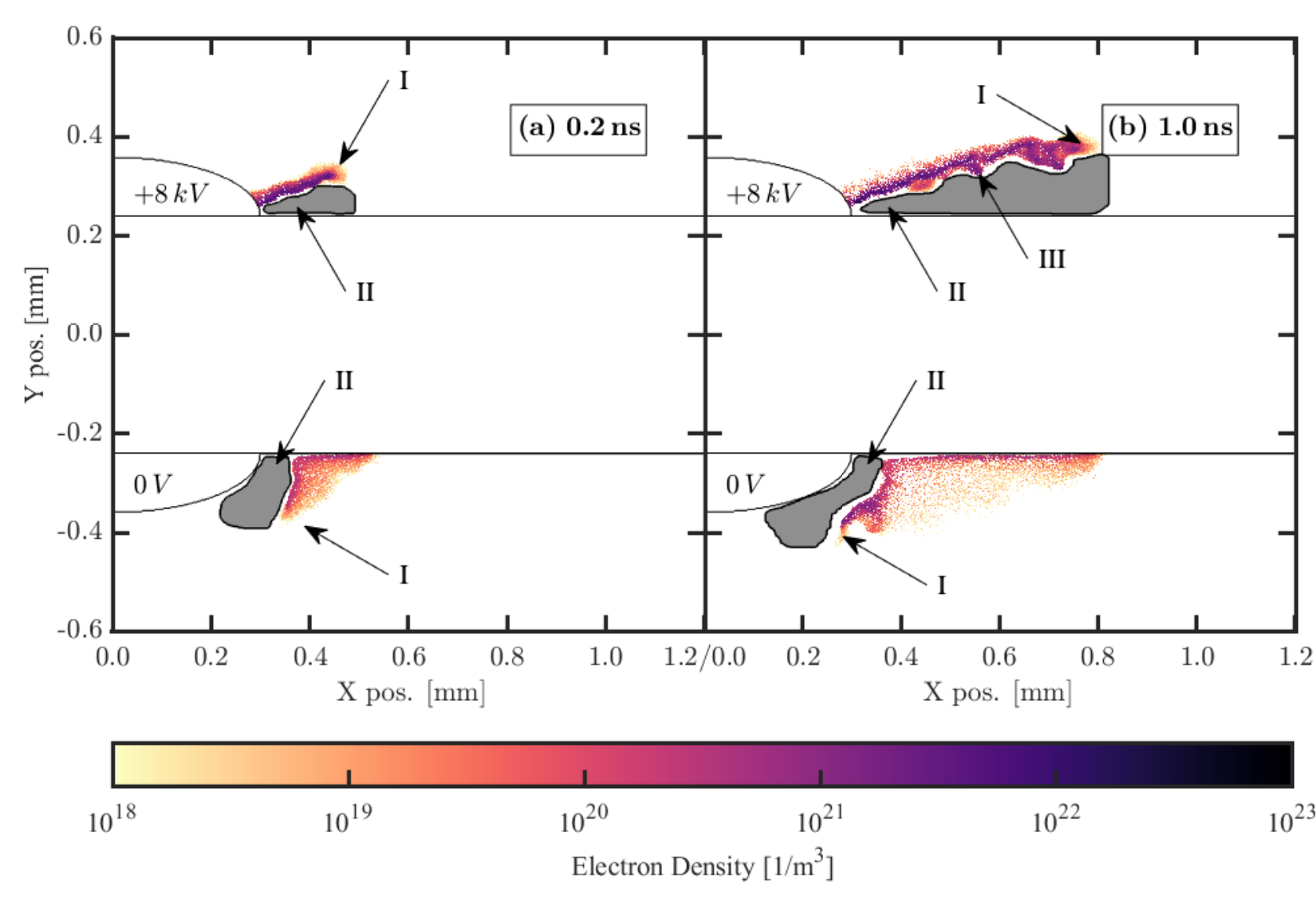}
            \caption{Spatial profiles of the electron density plotted on a logarithmic intensity scale of the dual streamer simulations with constant voltage. Sub figures (a) and (b) correspond to the timestamps of 0.2 and $1.0\,$ns, respectively. Features of importance are labeled, where the annotations are as follows: I) positively charged region/streamer head, II) location of electron depletion, i.e. sheath like feature., III) potential/failed positive streamer branch.}
            \label{fig:DC BothStreamers}
        \end{figure*}
        
        Newly created background electrons are pushed away from the cathode. Simultaneously, the electrons are attracted towards the positively charged region. Outside of the sheath region between the two, marked via an arrow labeled (II) in \cref{fig:DC NegativeStreamer}, these two directions are opposite one another. Only a very small amount of electrons are sufficiently accelerated to the positive charges with enough energy in order to cause ionization. Therefore, minimal propagation of the negative streamer parallel to the cathode surface takes place, as depicted via (I). Newly created background and avalanche electrons that reach the dielectric surface, instead of the positively charged spatial region, help to promote the propagation of the negative streamer along the surface of the dielectric in the X-direction away from the cathode and towards the center of the simulation area. However, no distinctly visible negatively charged streamer head is directly observable.
        
        At $0.2\,$ns the negative streamer has advanced $0.077\,$mm meaning a propagation speed of $0.39\,$mm/ns. By the end of the simulated time, $1.0\,$ns, the streamer had stopped propagating a significant amount. The negative streamer had reached a propagation distance of $0.25\,$mm resulting in an averaged speed of $0.25\,$mm/ns. The actual instantaneous speed of the streamer would be significantly slower at this timestamp, as the average includes the faster propagation of the early streamer. As with the positive streamer, lower and higher initial electron densities result in a shorter and longer propagation distance, respectively. Furthermore, under no conditions did the two simulated negative streamers next to both cathodes extend a significant amount into the X-direction, such that interactions between the two negative streamers do not need to be considered.
        
    \subsection{Dual Streamer Dynamics - DC}
    \label{DualStreamers}
        \begin{figure*}[t]
            \centering
            \includegraphics[width=0.885\textwidth]{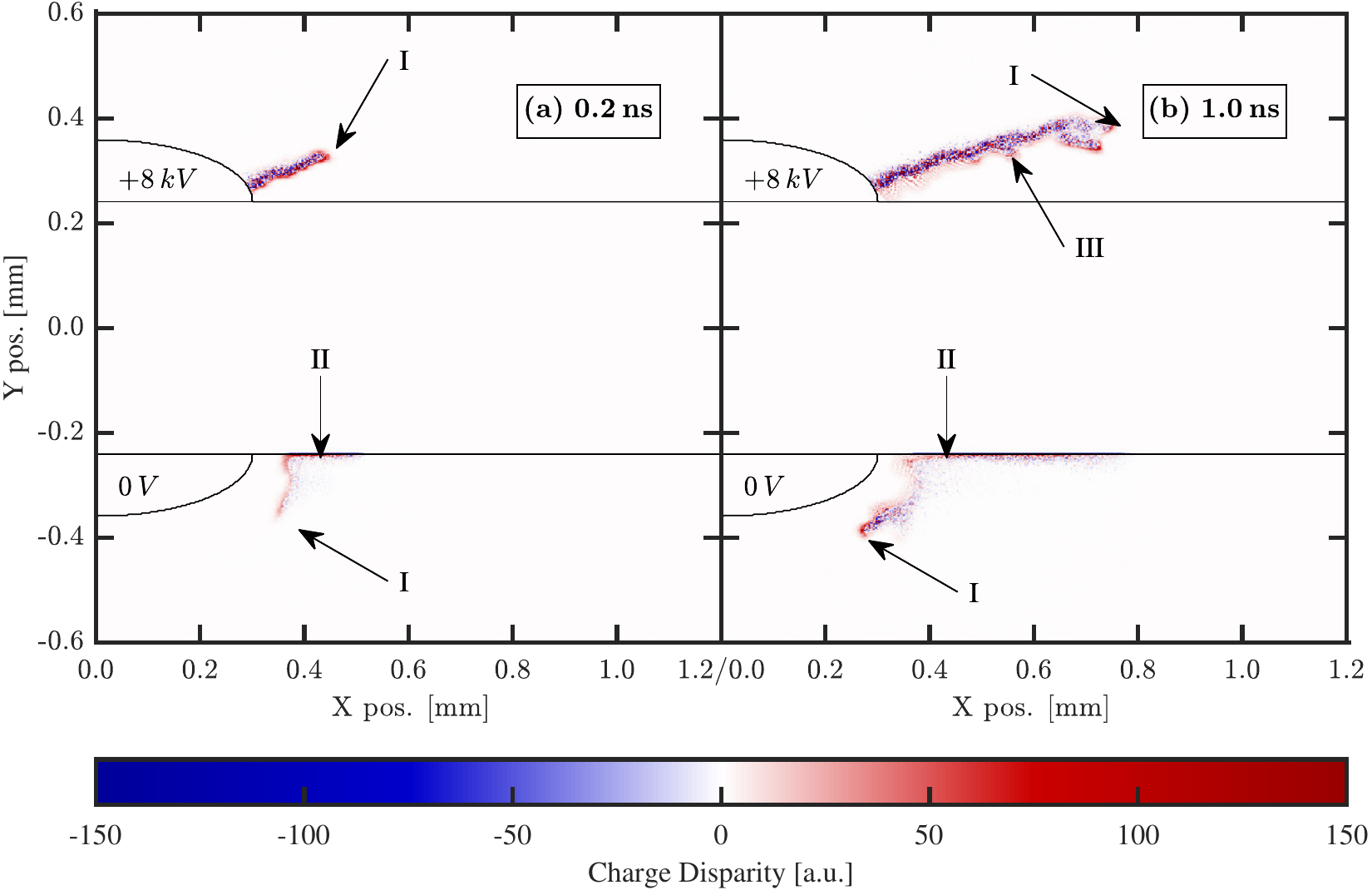}
            \caption{Spatial profiles of the charge disparity plotted on a diverging intensity scale of the dual streamer simulations with constant voltage. Sub figures (a) and (b) correspond to the timestamps of 0.2 and $1.0\,$ns, respectively. Features of importance are labeled with arrows, where the annotations are as follows: I) positively charged region/streamer head, II) surface charges which are visibly hidden by the mask of the dielectric barrier, III) potential/failed positive streamer branch.}
            \label{fig:DC BothStreamersCharge}
        \end{figure*}
    
        Presented in \cref{fig:DC BothStreamers,fig:DC BothStreamersCharge} is the complete DC scenario, where seed electrons are present on both the anodic and cathodic sides of the dielectric barrier. The same positive $8\,$kV DC voltage is used. Comparing \cref{fig:DC PositiveStreamer}(a), \cref{fig:DC NegativeStreamer}(a), and \cref{fig:DC BothStreamers}(a) a small difference is observed at $0.2\,$ns. Primarily, the sizes and overall density of both the positive and negative streamers have increased. The positive streamer has advanced $0.15\,$mm while the negative streamer has advanced $0.088\,$mm away from the anodes and cathodes, respectively. By $1.0\,$ns both streamers have significantly increased in size and average density compared to \cref{fig:DC PositiveStreamer}(b) and \cref{fig:DC NegativeStreamer}(b). Failed branches on the positive streamer are still present. The positive streamer has advanced a total of $0.41\,$mm while the negative streamer advanced a total of $0.27\,$mm. \Cref{tab:SizeAndSpeed} summarizes the streamer height, length, propagation angle, and propagation speed for the positive and negative streamers under all three simulation conditions. The propagation angle is determined as the angle at which the positive streamer propagates away from the dielectric surface, and is treated as $0\,^\circ$ for the negative streamer. The streamer length and thickness are respectively the size of the streamers with respect to the parallel and perpendicular axes about the streamer propagation angle.
        
        On the anodic side of the dielectric, the positively charged streamer head of the positive streamer is facing the dielectric surface, which can be seen as the red charges in \cref{fig:DC BothStreamersCharge}. This positively charged area acts as a virtual anode that leads to an enhanced electric field in both the X- and Y-directions below the dielectric surface on the cathodic side. Additionally, the positive streamer has a high charge density. The enhanced field and high density promote the expansion of the negative streamer along the surface of the dielectric in the X-direction. The negative streamer thus charges the surface of the dielectric even more. These negative surface charges along the dielectric barrier on the cathodic side act as a virtual cathode, enhancing the electric field in both the X- and Y-directions above the dielectric. Thus, the negative streamer also facilitates an easier expansion of the positive streamer in the X-direction. Here it is clear, that both streamers work together in a unison that increases the effective plasma surface coverage and volume of both streamers. Naturally, the electric field reduces proportional to the square of the distance from the electrodes, such that the positive and negative streamers are no longer able to expand any further, even with their cooperative effect being considered. Therefore, as with the single phase streamer simulations, interactions with the neighboring discharges on the right hand side of the simulation domain do not need to be taken into consideration.
        
        \begin{table*}[t]
            \centering
            \begin{tabular}{|c|l|c c|c c|c|c c|}
                \hline
                \multirow{2}{*}{Time} & \multirow{2}{*}{DC Streamer} & \multicolumn{2}{c|}{Thickness [$\mu$m]} & \multicolumn{2}{c|}{Length [$\mu$m]} & Angle [$^\circ$] & \multicolumn{2}{c|}{Speed [$\frac{\mu\mathrm{m}}{\mathrm {ns}}$]}\\
                 & & Average & Maximum & Average & Maximum & & Propagation & Lateral \\
                \hline \hline
                \rowcolor{gray!25}
                \cellcolor{white} & Positive & 38.39 & 49.20 & 123.4 & 170.4 & 20.60 & 617.1 & 577.7\\
                \rowcolor{white}
                \cellcolor{white} & Negative & 63.66 & 133.2 & 77.70 & 158.4 & -- & -- & 388.5\\
                \rowcolor{gray!25}
                \cellcolor{white} & Full (+) & 40.27 & 54.00 & 149.72 & 206.4 & 14.80 & 748.61 & 723.77\\
                \rowcolor{white}
                \cellcolor{white}\multirow{-4}{*}{$0.2\,$ns} & Full (-) & 64.63 & 128.4 & 88.20 & 166.8 & -- & -- & 441.0\\
                \hline
                \rowcolor{gray!25}
                \cellcolor{white} & Positive & 60.07 & 76.80 & 305.5 & 410.4 & 13.30 & 305.5 & 297.3\\
                \rowcolor{white}
                \cellcolor{white} & Negative & 79.10 & 115.2 & 247.9 & 372.0 & -- & -- & 247.9\\
                \rowcolor{gray!25}
                \cellcolor{white} & Full (+) & 65.92 & 80.40 & 409.17 & 516.0 & 10.50 & 409.17 & 402.31\\
                \rowcolor{white}
                \cellcolor{white}\multirow{-4}{*}{$1.0\,$ns} & Full (-) & 97.66 & 157.2 & 271.0 & 429.6 & -- & -- & 271.0\\
                \hline
            \end{tabular}
            \caption{Extracted average and maximum streamer heights and lengths of the DC streamer simulations at both output timestamps of $0.2\,$ns and $1.0\,$ns. The thickness and length are treated as the sum of cells perpendicular and parallel to the streamer propagation direction. The direction of the negative streamers is treated as parallel to the dielectric surface, while the angle of incidence of the positive streamers is determined in post analysis. The propagation speed is determined as the length of the streamer divided by the passed time. The lateral speed is the X-component of the propagation speed.}
            \label{tab:SizeAndSpeed}
        \end{table*}
        
        In essence, the positive streamer and the negative streamer work together to promote propagation. Both of the streamers are acting against the potential energy barrier of ionization and the ever decreasing electric field strength. Therefore, with the simultaneous ignition of both positive and negative streamers in a twin SDBD system, the surface coverage and plasma volume are significantly increased when compared to a submerged symmetric SDBD system. When comparing the average lengths of the single and dual streamers in \cref{tab:SizeAndSpeed}, the positive streamer sees an increase of the propagation length by $17.6 - 25.3\,\%$ and then negative streamer sees an increase of $8.5 - 11.9\,\%$ when both streamers are simultaneously ignited.

    \subsection{Dual Streamer Dynamics - AC}
    \label{ACStreamers}
\begin{figure*}[p]
            \centering
            \includegraphics[width=0.885\textwidth]{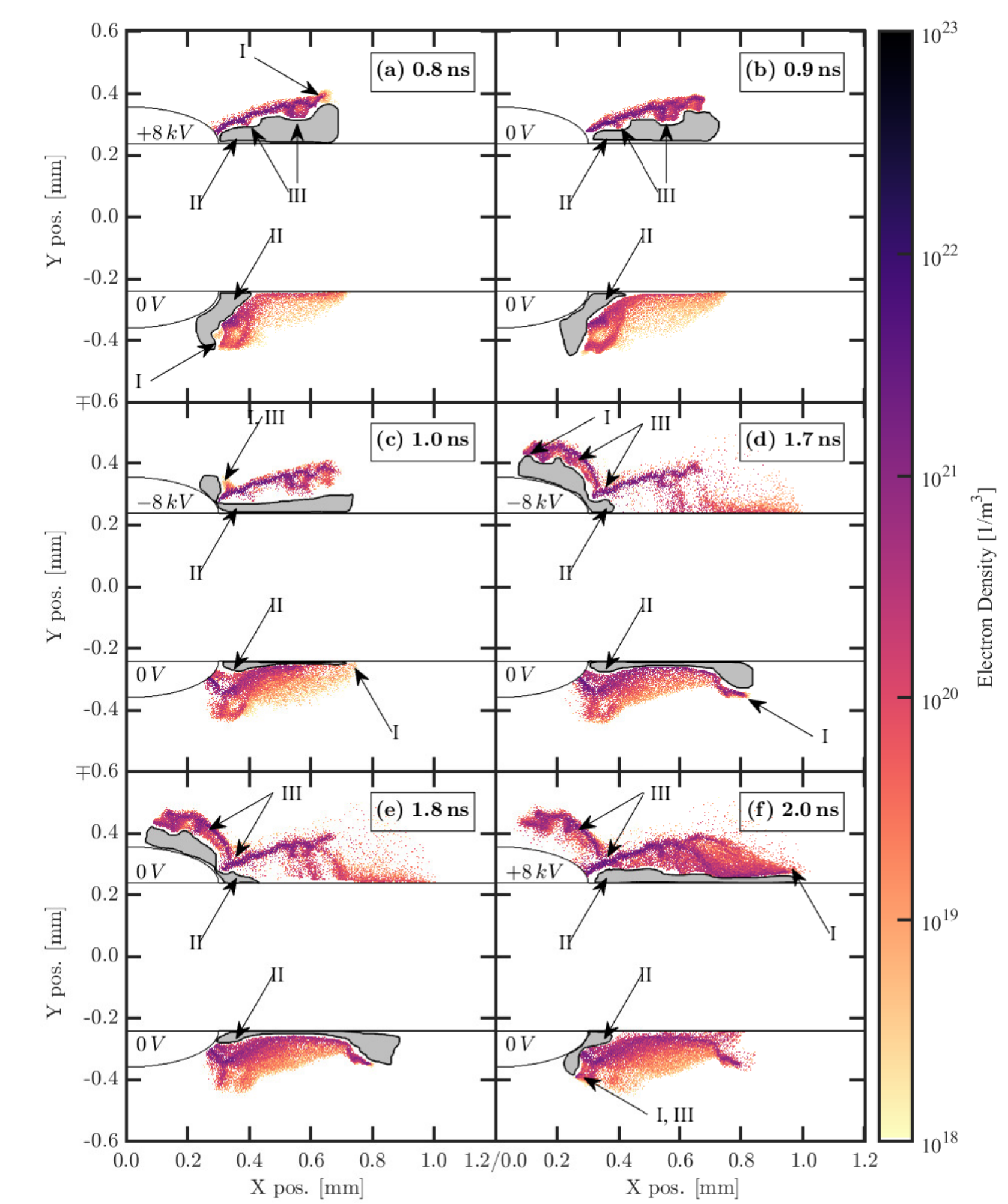}
            \caption{Spatial profiles of the electron density plotted on a logarithmic intensity scale at six chosen time stamps of the multi streamer simulations with switching voltage. Sub figures (a) through (f) correspond to the timestamps of 0.8, 0.9, 1.0, 1.7, 1.8, and 2.0$\,$ns, respectively. The applied voltages are respectively written within the electrode profiles. Features of importance are labeled with arrows, where the annotations are as follows: I) positively charged region leading to streamer propagation, II) shaded region showing location of electron depletion, \textit{i.e.} sheath like feature., III) potential/failed/completed positive streamer branch.}
            \label{fig:AC_Dens}
        \end{figure*}

\begin{figure*}[p]
            \centering
            \includegraphics[width=0.885\textwidth]{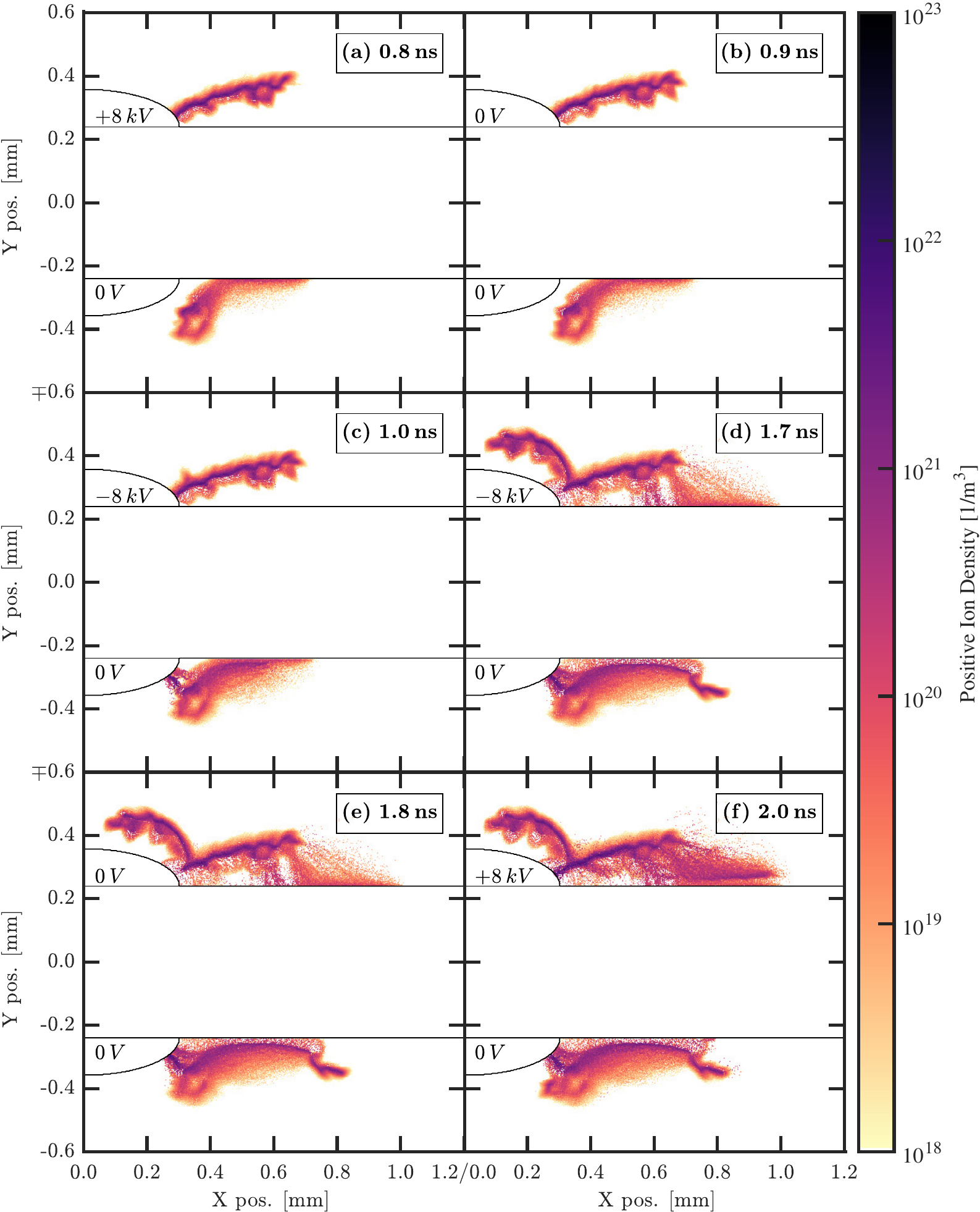}
            \caption{Spatial profiles of the positive ion density, \textit{i.e.} the summation of N$_2^+$ and O$_2^+$ ions, plotted on a logarithmic intensity scale at six chosen time stamps of the multi streamer simulations with switching voltage. Sub figures (a) through (f) correspond to the timestamps of 0.8, 0.9, 1.0, 1.7, 1.8, and 2.0$\,$ns, respectively. The applied voltages are respectively written within the electrode profiles.}
            \label{fig:AC_ionDens}
        \end{figure*}

\begin{figure*}[p]
            \centering
            \includegraphics[width=0.885\textwidth]{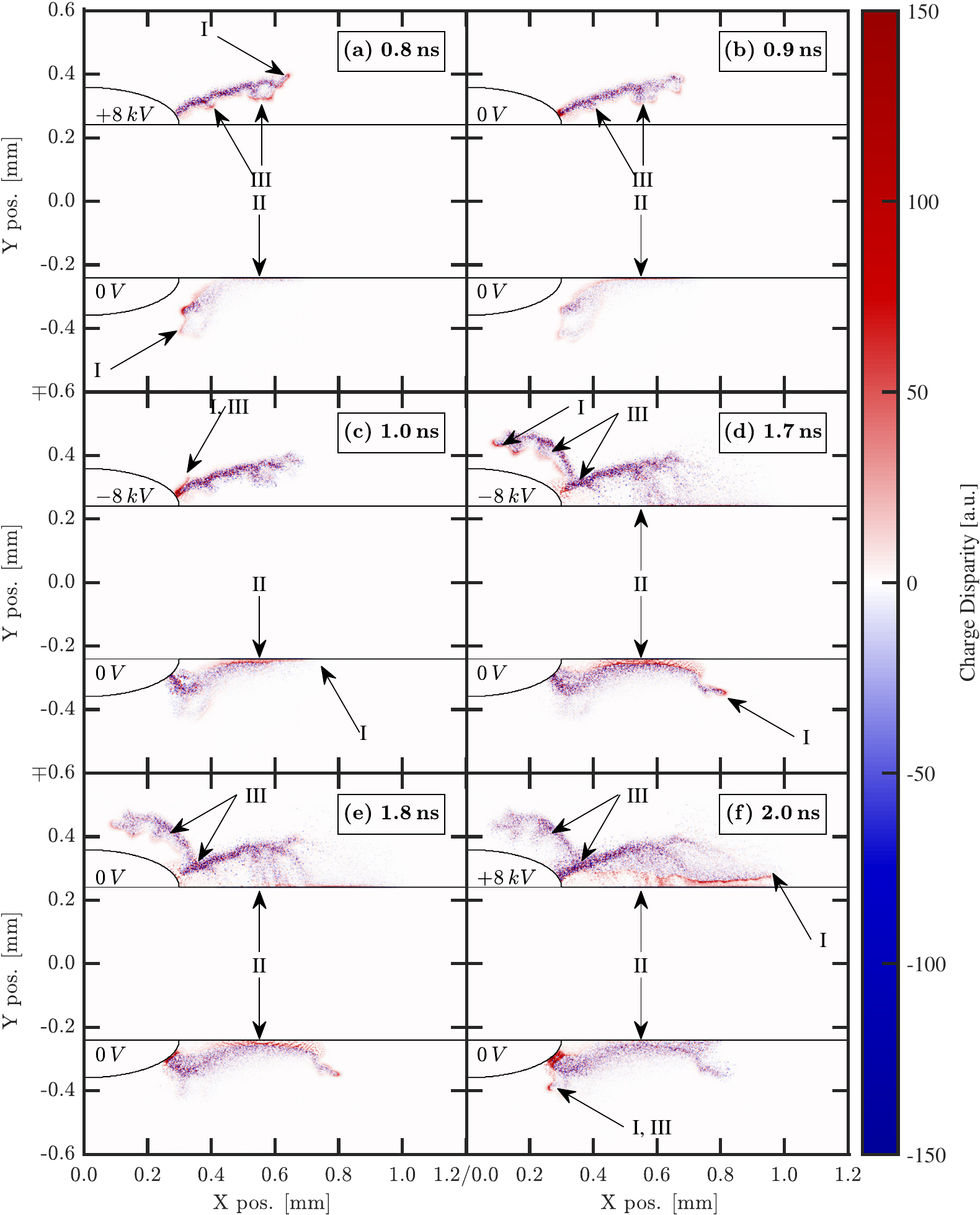}
            \caption{Spatial profiles of the charge disparity plotted on a diverging intensity scale at four chosen time stamps of the multi streamer simulations with switching voltage. Sub figures (a) through (f) correspond to the timestamps of 0.8, 0.9, 1.0, 1.7, 1.8, and 2.0$\,$ns, respectively. The applied voltages are respectively written within the electrode profiles. Features of importance are labeled with arrows, where the annotations are as follows: I) positively charged region leading to streamer propagation, II) surface charges which are visually hidden by the mask of the dielectric barrier, III) potential/failed/completed positive streamer branch.}
            \label{fig:AC_Charge}
        \end{figure*}

\begin{figure*}[p]
            \centering
            \includegraphics[width=0.885\textwidth]{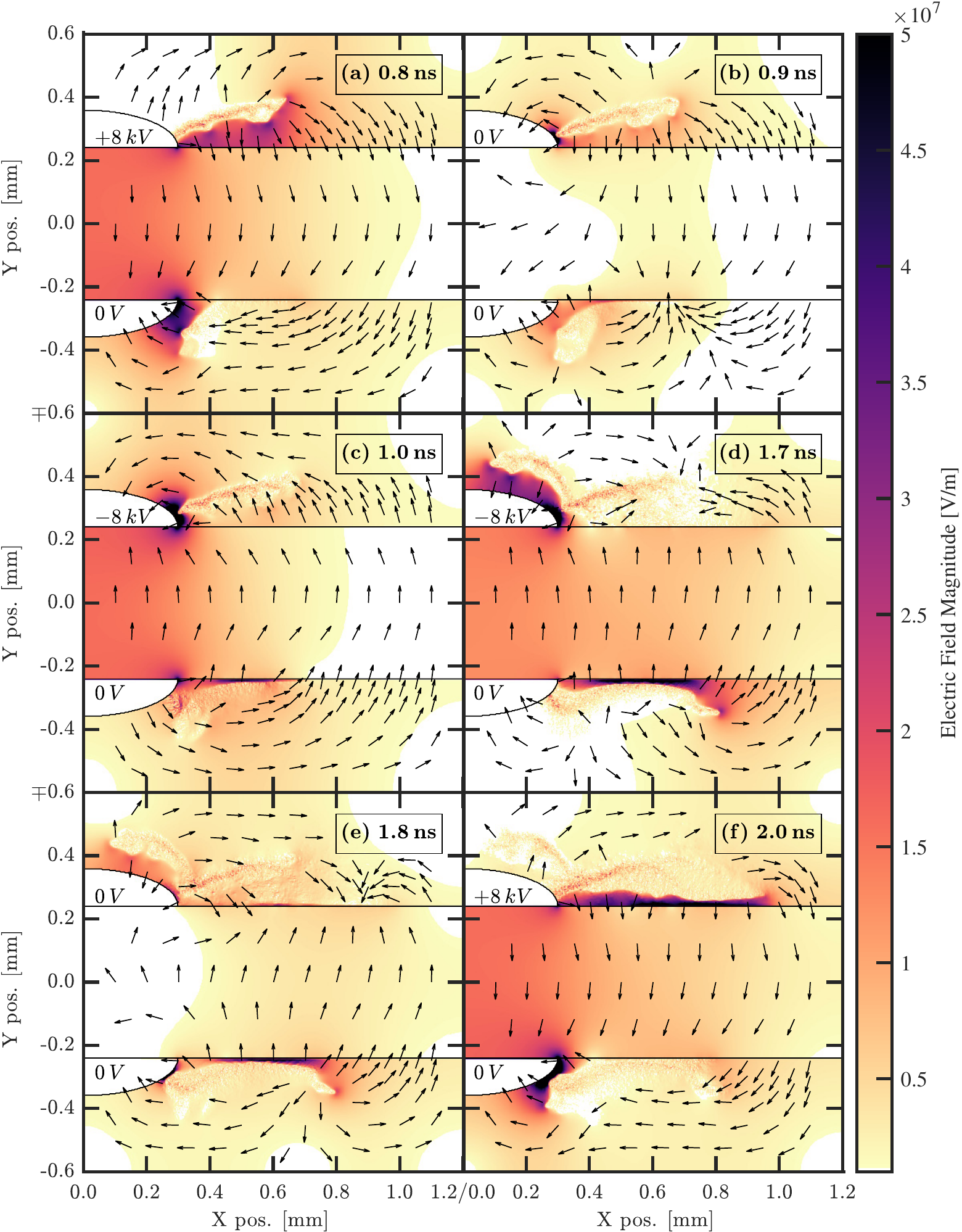}
            \caption{Spatial profiles of the absolute value of the electric field plotted on a linear intensity scale as well as directional arrows at four chosen time stamps of the multi streamer simulations with switching voltage. Sub figures (a) through (f) correspond to the timestamps of 0.8, 0.9, 1.0, 1.7, 1.8, and 2.0$\,$ns, respectively. The applied voltages are respectively written within the electrode profiles. Cut off value for minimum intensity scale (white) chosen as 1e6$\,$V/m. The direction of the electric field is shown via the normalized vector field as discussed in \cref{Waveform}}
            \label{fig:AC_EField}
        \end{figure*}

        Due to the minimal extension of the plasma into the free space above and below the dielectric surface of the simulations discussed in \cref{SingleStreamers,DualStreamers}, the simulated area was shifted horizontally to be centered about a single electrode pair, and reduced in width. Under this geometry, geometry(b), a bipolar AC square voltage profile with fast rise and short pulse times is simulated, shown in \cref{fig:ACPulseform}. Seed electrons are placed both above and below the dielectric barrier. Under such conditions, during the first positive pulse the plasma propagates near identically to the DC case discussed in \cref{DualStreamers,fig:DC BothStreamers,fig:DC BothStreamersCharge}. However, here it is observed that two near-mirror discharges simultaneously propagate about the horizontal center axis of both the anode and cathode. For reasons of consistency, only the right half of the simulated area is shown, as seen in \cref{fig:Geom(b)}. If shown, minimal differences between the left and right discharges would be seen, but may be attributed to the stochastic nature of the PIC/MCC code and the random seed electrons implemented each time step. Additionally, the implemented rising time of the voltage waveform from $0\,$V to $+8\,$kV at $0.1\,$ns does not contribute many differences, except perhaps a slightly reduced overall density and propagation distance. The electron density distribution, positive ion density distribution, \textit{i.e.} summation of N$_2^+$ and O$_2^+$ ions, charge disparity distribution, and electric field magnitude and direction are shown in \cref{fig:AC_Dens,fig:AC_ionDens,fig:AC_Charge,fig:AC_EField}, respectively. Sub-figures (a) through (f) of each correspond to identical timestamps of interest, shown with respect to the voltage waveform in \cref{fig:ACPulseform}.
        
        Between $0.8\,$ns and $1.0\,$ns the applied voltage is reduced, at $0.9\,$ns the applied voltage is $0\,$V, after which the role of the anode and cathode switch. Due to the polarity switch the electric field is reversed, thus the electrons move in the opposite directions. Free electrons present in the streamer above the dielectric move away from the now metallic cathode. Likewise, electrons from the  bulk of the streamer below the dielectric move towards the now anode. Electrons along the surface of the dielectric remain attached and do not move. At $1.0\,$ns the voltage on the cathode has reached its minimum value of $-8\,$kV, where it stays constant for a further $0.7\,$ns. After which, a second polarity switch takes place. All the while, the positive ion densities very closely follow the electron density profiles.

        \subsubsection{$1^{st}$ Polarity Shift - Positive to Negative streamer} \hfill\\
        
        Paying attention to the top half of the simulation regime focuses on the shift from a positive streamer to a negative streamer. As the voltage drops on the top electrode from $+8\,$kV to $0\,$V between $0.8\,$ns and $0.9\,$ns, sub-figures (a) and (b) respectively of \cref{fig:AC_Dens,fig:AC_ionDens,fig:AC_Charge,fig:AC_EField}, the electrons are not accelerated as strongly as before. The electrons relax and shift a little inwards towards the streamer bulk and the positively charged streamer head. The plasma volume slightly shrinks and the overall electron density becomes more refined and increases in number. The positively charged streamer head reduces in thickness and disparity, \textit{i.e.} becomes more quasi neutral. As the electrons are not as strongly/no longer attracted to the metallic anode, a positive space charge builds up at the streamer anchor on the anode. These two effects respectively lead to the electric field strength reducing between the streamer and the dielectric surface, and a very strong electric field between the anode and the streamer anchor. At $0.9\,$ns the quasi neutral streamer has a slight net positive charge, and thus takes on the role of a virtual anode along the boundaries of the streamer, meaning that the electric field above the streamer has reversed in the X-direction, but not the Y-direction.
        
        Between $0.9\,$ns and $1.0\,$ns, sub-figures (b) and (c) respectively, the applied voltage is negative, thus the metallic electrode is now the cathode and the dielectric surface is the anode. Due to the reversed electric field, electrons within the streamer begin falling to the dielectric surface. Along the way the remaining positive charges in the streamer head are flooded with electrons such that no charge disparity is noticeable, as it can be seen that the positive ion densities between the positive streamer and the dielectric do not change between these time steps. During this process, the electric field between the streamer and the dielectric surface completely reverses in both the X- and Y-directions. Naturally, falling electrons starting at locations where the streamer began to branch off but could not expand would reach the dielectric surface first. As the electrons are accelerated towards the dielectric surface, further ionization events take place creating new ions and electron avalanches. The electrons that first reach the dielectric charge the surface and repel other electrons into the X-direction away from the cathode, increasing the plasma propagation length. The original streamer is now acting like a negative streamer. By $1.7\,$ns, sub-figure (d), the negative streamer has charged the top surface of the dielectric and almost doubled the lateral length of the original positive streamer.
        
        Between $0.9\,$ns and $1.0\,$ns, the electrons near the streamer anchor/tail completely break away from the metallic cathode as the electrons are pushed away from it; however, the positive ions do not move. This results in a net positive charged being left behind. Thus a new positively charged streamer head forms between the cathode and the streamer bulk, both above and below the streamer. Along with this new streamer head forms an extremely high electric field in the local proximity, oriented away from the positive charges towards the cathode.
        
        Newly created electrons above the cathode and the streamer bulk, which is acting as an anode, are attracted to the streamer head and a small branch begins to form. This branch is shown in \cref{fig:AC_Dens,fig:AC_Charge}(c) with arrows labeled (III), and is also visible in \cref{fig:AC_ionDens}(c). As the simulation progresses in time, new electrons are continuously attracted towards this branch, gain energy, and eventually cause ionization. A cathode directed positively charged streamer head propagates along and floats above the cathode. A near mirror branch simultaneously forms on the other side of the metallic grid, which is not shown. Due to the positively charged streamer heads leading both of these branches, they repel one another. Therefore, neither branch is able to reach the other. By $1.7\,$ns, sub-figure (d), the branch has completely developed. Through multiple executions, it has been observed that this branching does not take place if the applied voltage, and thus electric field between the streamer head and cathode, is too low. It should be noted that the initial branching has a very similar structure to the positively charged spatial region of the negative streamers in \cref{fig:DC NegativeStreamer,fig:AC_Dens,fig:AC_ionDens,fig:AC_Charge}(a), and \cref{fig:DC BothStreamers}(b). One could expect that given a high enough voltage, the positive space charges would continue to wrap around the cathode in the same manner as the branching in \cref{fig:AC_Dens,fig:AC_Charge}(c) and (d). Therefore, the branching should not be considered as solely limited to the polarity switches, but rather that they are encouraged by the polarity switches. As with the positive streamer in the DC case, discussed in \cref{SingleStreamers}, the authors believe these simulated branching mechanisms are accurate, even given the difficulty of experimentally observing them.
        
        \subsubsection{$1^{st}$ Polarity Shift - Negative to Positive Streamer} \hfill\\

        Focusing now on the bottom half of the simulation regime tracks the shift of the negative streamer to a positive streamer between $0.8\,$ns and $1.0\,$ns. During this time, the applied voltage is switched from $+8\,$kV to $-8\,$kV; however, the bottom electrode is held at a constant $0\,$V. As the applied voltage changes polarity, the bottom electrode also switches roles, now becoming the anode. Unlike the top half, the relaxation of the electric field causes a small shift in the bulk electrons which leads to both a large increase in the streamer size as the electrons are pushed away from the dielectric and towards the metallic electrode. Similar to the top half, the average electron density slightly increases and the charge disparity in the positively charged streamer head near the now anode reduces. This eventually leads to the streamer attaching to the anode, seen in in sub-figure (c), as electrons are freely absorbed by it. This motion also leads to the creation of a strong positive ion density within at the anchor position, as seen in in \cref{fig:AC_ionDens}(c).
        
        Furthermore, a small positive space charge forms between the negatively charged dielectric surface and the bulk plasma as the electrons are pushed away from the dielectric surface, but the positive ions do not move. However, the electrons that had attached to the surface do not desorb within the simulation, neither are electrons emitted due to surface field emission, emitted due to ion induced secondary electrons, nor are electrons reflected. The newly formed positively charged head and the negative surface charges form a very high electric field in a very thin sheath like structure between the streamer and the dielectric surface by $1.0\,$ns. The positively charged streamer head is floating above the surface, which is acting as the cathode; however, due to the original proximity of the bulk plasma to the surface and the surface charges, the streamer head remains very close to the surface.
        
        The proximity of the streamer head limits the ability of the streamer to propagate into the X-direction. As electrons are continuously pushed away from the dielectric surface, the thickness of the streamer head and consequentially the sheath like region increase. Eventually, near the "tip" of this region, along the X-direction, newly generated electrons outside of the plasma bulk are sufficiently attracted towards the positive charges. This leads to the streamer head curling around the "tip" of the streamer bulk, providing a virtual anode for further newly created electrons to be attracted to. Sufficiently energetic electrons will promote propagation further into the X-direction, extending the plasma. This propagation also significantly extends in the Y-direction away from the dielectric surface as electrons created near the surface will not gain enough energy for ionization. This causes the streamer to properly float above the dielectric surface, which can be seen at $1.7\,$ns in sub-figure (d), as expected of a cathode directed positively charged streamer head. The increased propagation length is not as significant as the streamer on the top half of the dielectric, due to the limiting effect that the surface streamer exhibited. If the surface of the dielectric was not considered as a pure absorber, then the emission and reflection features would provide an additional electron source that would promote the expansion and propagation of the streamer after the voltage had switched.
        
        \subsubsection{$2^{nd}$ Polarity Shift} \hfill\\
        
        Between $1.7\,$ns and $1.9\,$ns, the applied voltage potential begins to switch again, this time rising from $-8\,$kV to $+8\,$kV. At $1.8\,$ns, the second polarity change occurs. Due to limited computational resources, the simulation was not executed for a second full positive cycle, and was instead ended at $2.0\,$ns. During this polarity switch, the same changes in the positive and negative streamers are observed.
        
        On the bottom half of the simulation, the shift from a positive streamer at $1.7\,$ns, sub-figures (d) to a negative streamer is observed. When the applied voltage is $0\,V$ at $1.8\,$ns, sub-figures (e), it can be seen that the floating positively charged streamer head is beginning to be flooded, while a new positively charged streamer head is forming near the metallic cathode. By $2.0\,$ns, sub-figures (f), the streamer bulk has mostly reached the dielectric surface again, has expanded further in the X-direction, and a new positive streamer branch forms near the cathode. It is very well expected that this branch would behave as the one discussed above.
        
        On the top half of the simulation, not only is the shift from a negative to positive streamer observed, but also the beginning of the collapse of the positive streamer branch is observed. As already explained and expected, at $1.8\,$ns the positively charged streamer head of both the negative streamer and the streamer branch near the metallic electrode is flooded by electrons moving towards the new anode. At $2.0\,$ns the anchor of the main streamer on the anode is fully formed; however, the electrons within the branch have a further distance to travel and have not yet reached the anode. As the polarity switches at $1.8\,$ns, electrons near the dielectric surface are repelled away and a floating positively charged streamer head forms. At $2.0\,$ns this streamer head is beginning to wrap around the large streamer bulk to promote further expansion in both the X- and Y-directions away from the metallic anode and dielectric surface respectively.
        
        \subsubsection{Summary of Polarity Shifts} \hfill\\
        
        During both polarity shifts, similar and important events take place on the respective positive and negative streamers. The Negative streamer is initially attached to the anodic dielectric surface, and floating away from the metallic cathode. As the polarity changes, the electrons reverse in direction, attaching to the metallic anode and forming a positively charged streamer head near the dielectric surface. Newly created electrons are quickly attracted to the streamer head and as such allow for the now positive streamer to further propagate into the X- and Y-directions, thereby increasing the volume and overall density of the streamer. The positive streamer is initially floating away from the cathodic dielectric surface, and attached to the metallic anode. As the polarity changes, electron avalanches are instigated and rush towards the dielectric surface, thereby drastically increasing the plasma density, volume, propagation length, and surface coverage. Additionally, as a positively charged streamer head and sheath like region form near the metallic cathode, newly created electrons are able to instigate an additional positive streamer branch that floats above the metallic cathode. This branching feature also drastically increased the electron density and volume. Given a high enough initial voltage, it is expected that this positive streamer branch could form on the negative streamer before any polarity switching occurs.
        
        The increase in plasma densities, volume, and surface coverage are expected to be directly beneficial to various applications such as plasma enhanced catalysis and gas treatment. In plasma enhanced catalysis, the dielectric surface will typically be coated with a catalyst, such that any increase in surface coverage directly increases the active area of the catalyst. Additionally, any increase in plasma volume and density will naturally increase the radical densities which are available to react with either the catalytic surface and or the treatment gas that the plasma is ignited in, thus directly affecting the efficiency of the process.

\section{Conclusions and Outlook}
\label{Conclusion}
    In this work, the plasma streamer propagation of a twin SDBD setup by means of PIC/MCC code modeled in dry air under DC and AC voltage operation. The AC driving voltage waveform corresponded to a nanosecond square waveform with sub-nanosecond risetimes. The twin SDBD geometry being fully exposed and symmetric about the dielectric layer promotes both positive and negative streamer discharges to ignite simultaneously, along the edges of both the anode and cathode. This symmetry has not been theoretically investigated extensively, leaving the question of, among others, how do the streamers affect one another. In order to provide insight into this question, multiple scenarios were simulated. First, the propagation of a positive streamer and negative streamer were simulated individually under identical DC conditions. Second, both streamers were allowed to propagate using the same DC conditions, thereby providing insight into the interplay of the two streamers. However, the main focus of the paper is on the role of how the streamers interact and change under AC conditions; therefore, a short multi nanosecond duration bipolar square pulse is used to approximate said conditions.
    
    It was first shown that both the positive and negative streamers behave as expected under DC conditions. Both streamers form a quasi neutral bulk. The positive streamer forms and propagates via a floating cathode directed positive streamer head, while the negative streamer propagates via an electron avalanche along the surface of the dielectric barrier. The negative streamer also forms a positive space charge that floats above the metallic cathode. The floating positive space charges of both the positive and negative streamer must float as new electrons which are introduced between the cathode and said space charges are not able to gain enough energy for new ionization events. It was then shown that the interaction of both streamers under DC conditions does not significantly alter the propagation methods, but that the positive streamer "pulls" the negative streamer while simultaneously being "pushed" by the negative streamer, effectively increasing the surface coverage and the densities of the plasma streamers. The speed of propagation of both streamers differs when individually simulated versus when simultaneously simulated. The positively charged streamer head of the positive streamer propagates away from the anode providing an enhanced electric fields that the negative charges of the avalanche of the negative streamer then follow. Likewise, the negative streamer charges the dielectric surface which then helps to push the positively charged streamer head of the positive streamer further away from the anode. 
    
    Next, the interactions of the two streamers under switching voltage conditions was investigated. The fast polarity switching of the applied voltage causes significant changes in the streamers. The switch from a positive streamer to a negative streamer, and vice versa were observed to cause a significant increase in both plasma size and density due to similar effects that take place during the DC scenario. It was also observed that additional positive streamer branches are able to form between the negative streamer and cathode under the given conditions. The initial branching structure is very similar to structures that formed on the negative streamer during DC conditions and the AC conditions before any voltage switches. Therefore it is hypothesized that the voltage switching allows for a branch to more easily form, but is still subject to some minimal necessary applied voltage for a given set of geometrical conditions.
    
    Overall, an electrode geometry allowing for two oppositely-phased plasmas to simultaneously ignite is beneficial with respect to plasma size and density. The two fully exposed electrodes create strongly curled electric fields that promote the ignition of plasma streamers near the surface of the dielectric. The simultaneous ignition of the streamers enhances the lateral electric fields causing the streamers to propagate further away from the metallic electrodes than they would if one electrode was submerged. This effect is even further enhanced if the applied voltage is able to quickly switch polarities before the streamers have a chance to self extinguish; however, this fast of a voltage profile is difficult to experimentally achieve and as such the reader should remember this if attempting to compare any numerical information from this paper. Nonetheless, the enhanced electric fields also allow the plasma to achieve higher densities, which is in many applications desirable.

    In plasma enhanced catalysis applications, one might want to coat the dielectric surface with a catalyst. Having an enhanced plasma propagation length would directly correlate to an increased surface area of the catalyst that is directly affected by the plasma, leading to a potentially enhanced efficiency. In gas treatment applications, an increased plasma density is typically desirable in order to increase the rate of molecular fragmentation and/or purification. Future experimental measurements and theoretical and or numerical investigations on the electrode geometry could optimize an electrode system for a given set of applications. Additional simulations of a porous catalytic coating attached to the dielectric surface would provide additional insight into plasma enhanced catalysis applications. 
    
\section{Acknowledgements}
      This work is supported by the German Research Foundation (DFG) with the Collaborative Research Centre CRC1316 projects A4 and A5 and the Scientific Research Foundation from Dalian University of Technology, DUT19RC(3)045, and the National Science Foundation of China Grant No. 12020101005.

\color{black}
\newpage
\section*{ORCID iDs}
\begin{table}[h]
    %\centering
    \begin{tabular}{l p{4cm}}
        Q. Z. Zhang: & \url{https://orcid.org/0000-0002-5726-0829} \\
        R. T. Nguyen-Smith: & \url{https://orcid.org/0000-0002-5755-4595}\\
        F. Beckfeld: & \url{https://orcid.org/0000-0001-8605-2634}\\
        Y. Liu: & \url{https://orcid.org/0000-0002-2680-1338}\\
        T. Mussenbrock: & \url{http://orcid.org/0000-0001-6445-4990} \\
        J. Schulze: & \url{https://orcid.org/0000-0001-7929-5734}\\
        
    \end{tabular}
    \label{tab:my_label}
\end{table}

%\section{References}
\printbibliography
\end{document}